\documentstyle[12pt,aaspp4]{article}

\def\sn{SN~1987A}
\def\IUE{{\it IUE~}}
\def\HST{{\it HST~}}
\def\Ha{{\rm H}\alpha}
\def\Hb{{\rm H}\beta}
\def\Msun{M_\odot}
\def\EE#1{\times 10^{\small#1}}
\def\cm3{\rm ~cm^{\small -3}}
\def\kms{\rm ~km~s^{\small -1}}
\def\ergs{\rm\,erg~s^{\small -1}}
\def\ergcms{\rm\,erg~cm^{\small -2}~s^{\small -1}}

\def\EBV{{\rm E}(B-V)}
\def\wl{$\lambda$}
\def\wll{$\lambda\lambda$}

\def\HII{{\ion{H}{2}\,}}
\def\HeI{{\ion{He}{1}\,}}
\def\HeII{{\ion{He}{2}\,}}
\def\HeIII{{\ion{He}{3}\,}}
\def\CII{{\ion{C}{2}\,}}
\def\CIII{{\ion{C}{3}\,}}
\def\CIIIb{{\ion{C}{3}]\,}}
\def\CIV{{\ion{C}{4} }}

\def\NIIb{{\ion{N}{2}]\,}}
\def\NIII{{\ion{N}{3}\,}}
\def\NIIIb{{\ion{N}{3}]\,}}
\def\NIV{{\ion{N}{4}}}
\def\NIVb{{\ion{N}{4}]\,}}
\def\NV{{\ion{N}{5}\,}}
\def\NVI{{\ion{N}{6}\, }}
\def\OIIb{{\ion{O}{2}]\,}}
\def\OIII{{\ion{O}{3}\,}}
\def\OIIIb{{\ion{O}{3}]\,}}
\def\OIVb{{\ion{O}{4}]\,}}
\def\AlIII{{\ion{Al}{3} }}

\def\SIIb{{\ion{S}{2}]\,}}
\def\SiII{{\ion{Si}{2} }}
\def\SiIII{{\ion{Si}{3} }}
\def\SiIV{{\ion{Si}{4} }}
\def\FeII{{\ion{Fe}{2} }}

\begin{document}
\textwidth=6.9in

\title{The Evolution of Ultraviolet Emission Lines From \\
    Circumstellar Material Surrounding \sn}

\author{George Sonneborn\altaffilmark{1,2}, Claes 
Fransson\altaffilmark{3},
 Peter Lundqvist\altaffilmark{3}, Angelo Cassatella\altaffilmark{4}, \\
Roberto Gilmozzi\altaffilmark{5}, Robert P. Kirshner\altaffilmark{6}, 
Nino Panagia\altaffilmark{7,8}, \\
and Willem Wamsteker\altaffilmark{9}}
\affil{~ }


\altaffiltext{1}{Laboratory for Astronomy and Solar Physics, Code 681, NASA/Goddard Space Flight Center, Greenbelt, MD 20771; ~~sonneborn@fornax.gsfc.nasa.gov}
\altaffiltext{2}{Institut d'Astrophysique de Paris, 98bis boulevard Arago, 75014 Paris, France}
\altaffiltext{3}{Stockholm Observatory, S-133~36 Saltsj{\"o}baden, Sweden;~~claes, peter@astro.su.se}
\altaffiltext{4}{Istituto di Astrofisica Spaziale, CNR, CP 67, I-00044 Frascati, Italy;~~ac@vilspa.esa.es}
\altaffiltext{5}{European Southern Observatory, Karl-Schwarzschild Str.~2, D-85748 Garching bei M\"unchen, Germany;~~rgilmozz@eso.org}
\altaffiltext{6}{Center for Astrophysics, 60 Garden Street, Cambridge, MA 02138;~~kirshner@cfa.harvard.edu}
\altaffiltext{7}{Space Telescope Science Institute, 3700 San Martin Dr., Baltimore, MD 21218;~~panagia@stsci.edu}
\altaffiltext{8}{Affiliated with the Astrophysics Division, Space Sciences Department of ESA}
\altaffiltext{9}{\IUE Observatory, ESA-VILSPA, Casilla 50727, E-28080 Madrid, Spain;~~ww@vilspa.esa.es}

\begin{abstract}
The presence of narrow high-temperature emission lines from 
nitrogen-rich gas close to \sn\ has been a principal observational 
constraint on the evolutionary status of the supernova's progenitor.  
A new analysis of the complete five-year set of low and high
resolution \IUE ultraviolet spectra of \sn\ (1987.2--1992.3) 
provide fluxes for the \NV\wl1240, \NIVb\wl1486, \HeII\wl1640, 
\OIIIb\wl1665, \NIIIb\wl1751, and \CIIIb\wl1908 lines with 
significantly reduced random and systematic errors and reveals 
significant short-term fluctuations in the light curves. The \NV, 
\NIVb, and \NIIIb\ lines turn on sequentially over 15 to 20 days and 
show a progression from high to low ionization potential, implying an 
ionization gradient in the emitting region.  The line emission turns 
on suddenly at $83 \pm 4$ days after the explosion, as defined by 
\NIVb.  The \NIIIb\ line reaches peak luminosity at $399 \pm 15$ days.  
A ring radius of $(6.24\pm0.20)\EE{17}\rm\,cm$ and inclination of 
$41\fdg0 \pm 3\fdg9$ is derived from these times, assuming a circular 
ring.  The probable role of resonant scattering in the \NV\ light 
curve introduces systematic errors that leads us to exclude this line 
from the timing analysis.  A new nebular analysis yields improved CNO 
abundance ratios of N/C$=6.1\pm1.1$ and N/O$=1.7\pm0.5$, confirming 
the nitrogen enrichment found in our previous paper.  From the late-time 
behavior of the light curves we find that the emission originates 
from progressively lower density gas and that the emitting region has a 
multi-component density structure.  We estimate 
the emitting mass near maximum $(\sim 400$ days) to be $\sim 4.7\EE{-2} \Msun$, 
assuming a filling factor of unity and an electron density of $2.6\EE4 
\cm3$.  These results are discussed in the context of current models 
for the emission and hydrodynamics of the ring.

\end{abstract}


\section{INTRODUCTION }
 
Narrow emission lines from \sn\ were first seen with the {\it International
Ultraviolet Explorer} (\IUE) satellite in 1987 May (Fransson et al. 1989,
hereafter Paper I). The lines detected were \NV\wll1239, 1243, 
\NIVb\wl1486, \HeII\wl1640, \OIIIb\wll1661, 1667, \NIIIb\wll1747-1754, and
\CIIIb\wll1907, 1909. The absence of the usually strong \SiIV\wl1400 and
\CIV\wl1550 resonance lines can be attributed to absorption of these
transitions by the same ions in hot interstellar gas in the LMC. The presence
of \NV\ in the \sn\ spectra, on the other hand, is in fact an indication
that there is a negligible column density of interstellar \NV\ in the LMC. A
remarkable feature of the early observations of \sn\ was that the \NIIIb,
\NV, and \CIIIb\ lines increased in flux linearly with time. The \HeII, \NIVb,
and \OIIIb\ lines were weak and it was difficult to say anything conclusive
about their light curves. The increase in the fluxes stopped just after the
last observations discussed in Paper I, $\sim 400$ days after
explosion. Subsequent observations showed that the \CIIIb, \NIIIb, and 
\NV\ lines reached a maximum $\sim405$ days after the explosion (Sonneborn
et al. 1988). After this epoch the observations reported in this paper show a
steady decay of the lines. Earlier versions of the \IUE emission line 
light curves, but not the new measurements discussed in this paper, have 
been presented by us in several papers (Sonneborn et al. 1990; Panagia 
et al. 1991; Sonneborn 1991) and an independent analysis by Sanz 
Fern\'andez de C\'ordoba (1993). Preliminary versions of some of the 
new light curves have been shown by Fransson \& Sonneborn (1994) and 
Plait et al.  (1995).  In a complementary paper to this one, Pun et 
al.  (1995) analyzed the UV spectrophotometric evolution of the SN\,1987A 
debris from day 1.6 to 1567 in \IUE spectra covering 1150-3300\AA.  
They found good agreement with UV continuum observations of SN\,1987A 
with {\it Hubble Space Telescope} (\HST) and ground-based photometry where they 
overlap.

The presence and temporal development of these highly ionized lines
have been explained as a result of recombination, 
cooling and light echo effects (Lundqvist \& Fransson 1987; Chevalier 1988; 
Paper I; Lundqvist \& 
Fransson 1991, henceforth LF91).  The initial 
pulse of soft X-rays accompanying the breakout of the shock wave ionizes 
the circumstellar gas expelled by the progenitor star.  As the outburst 
evolves an increasingly large volume of the ionized gas becomes visible 
because of light travel-time effects.  If the recombination time is long 
compared to the epoch of observation, the observed flux is expected to grow 
in proportion to the emitting volume within the light echo paraboloid.  For 
a non-recombining ion emitting at constant temperature this leads to a flux 
increasing monotonically with time, until the whole structure is within the 
light echo paraboloid.  In reality, the temperature and ionization decrease 
with time, causing the evolution to become highly complicated.  Detailed 
modelling of the early line emission showed that the peak radiation 
temperature at the time of the breakout was between $4\EE{5} \rm \,K$ and 
$8\EE{5}\rm\,K$ (Fransson \& Lundqvist 1989; Lundqvist 1991).  A 
subsequent analysis, taking the ring geometry into account, resulted in a 
peak radiation temperature in excess of $10^6\rm\,K$ (Lundqvist \& Fransson 
1996, henceforth LF96).  These emission line observations therefore 
provide 
important information about the first moments of the supernova explosion.

Another important piece of information was the large nitrogen enrichment 
derived from the emission line ratios, primarily the \NIIIb/\CIIIb\ and 
\NIIIb/\OIIIb.  Abundance ratios of N/C $= 7.8 \pm 4$ and N/O $= 1.6 \pm 
0.8$, by number, found in Paper I strongly suggest CNO enriched gas.  This 
has proven to be one of the main observational constraints on the 
progenitor evolution (e.g., Saio, Nomoto, \& Kato 1988; Woosley 1988; Weiss 
1991; Podsiadlowski 1992), implying that the progenitor was in a post 
He-core burning phase at the time of the explosion.

After the discovery of the UV emission lines, circumstellar emission 
has also been observed at other wavelengths.  In the optical a large 
number of lines from both medium ionized species and low ionization 
ions have been detected (Allen, Meikle, \& Spyromilio 1989; Wampler \& 
Richichi 1989; Wampler, Richichi, \& Baade 1989; Crotts \& Kunkel 
1991; Kahn \& Duerbeck 1991; Meikle et al.  1991; Menzies 1991; Wang 
1991; Cumming 1994).  Extended loop-like structures surrounding 
SN\,1987A were first imaged in H$\alpha$+[\NIIb and [\OIIIb emission 
by Crotts, Kunkel, \& McCarthy (1989).  Even in these early images of 
the inner emission region appeared to be a hollow oval.  These 
structures were also seen in optical continuum, indicating the 
presence of time-delayed reflections by dust.  Imaging with the New 
Technology Telescope resolved the inner emission region of radius 
$\sim1\arcsec$ and several outer structures (Wampler et al.  1990; 
Wang \& Wampler 1992).  \HST images in [\OIIIb\wll4959, 5007 (Jakobsen 
et al.  1991, 1994; Plait et al.  1995) revealed the now well-known 
thin ring geometry of the inner region, rather than a spherical shell 
as had been previously hypothesized.  Further \HST observations showed 
that the outer structures are concentrated mainly to two off-center 
rings, which extend $\sim 2\farcs5$ to the north and south of the 
supernova (Burrows et al.  1995).  Crotts, Kunkel, \& Heathcote (1995) 
show that the toroidal geometry extends also to regions of the 
circumstellar medium exterior to the ring, and that there may be some 
connection between the inner and outer rings.  They also find that the 
inner ring is circular to a high degree of accuracy.  In addition to 
the geometry, the dynamics and origin of the inner ring is severely 
constrained by the emission line widths.  In Paper I we showed that 
the \CIIIb\wll1907, 1909 lines were very narrow $(FWHM<30\kms)$ and 
unshifted with respect to the LMC ISM toward the supernova $(v=285\pm 
6\kms)$.  Crotts \& Heathcote (1991) and Cumming et al.  (1996) found 
that the ring expands with a velocity of $10.3 \pm 0.4 \kms$.  They, 
as well as Wood \& Faulkner (1987) and Meikle et al.  (1991), also 
find a velocity gradient over the nebula, which is a strong indication 
of a ring geometry rather than a spherical shell.

Prompt radio emission from \sn, the result of shock interaction with 
the progenitor's stellar wind close to the photosphere, was observed during 
the first few days after explosion (Turtle et al. 1987).  This emission 
decayed away after the first week.  However, around day 1200 there was a 
revival of the radio emission and a steady increase in the flux was 
observed (Staveley-Smith et al.  1992, 1993) which continued through day 
2750 (Ball et al.  1995).  Chevalier (1992) and Duffy, Ball, \& Kirk (1995) 
attribute the resurgence of the radio emission from \sn\ to the 
interaction of the supernova ejecta with the density enhancement expected 
at the position of the termination shock in the blue supergiant wind of the 
progenitor.  Further evidence of circumstellar interaction is the 
increasing late-time 
X-ray emission observed by ROSAT (Beuermann, Brandt, \& Pietsch 1994; 
Gorenstein, Hughes, \& Tucker 1994; Hasinger, Aschenbach, \& 
Tr\"umper 1996).  Chevalier \& Dwarkadas (1995) point 
out that the density of the shocked blue supergiant wind is insufficient to 
explain the late radio and X-ray fluxes, as well as the apparent slowing 
down of the supernova shock (Staveley-Smith et al.  1993).  They propose 
that the supernova shock is now moving into the denser ($n_e \sim 10^2 
\cm3$) shocked red supergiant wind which was ionized by the progenitor.

In this paper we discuss the complete set of UV emission lines in \sn\ 
derived from \IUE low and high dispersion SWP spectra from the explosion up 
to 1992 April 20 (day 1882).  A new method of analysis has allowed us to 
reduce systematic and random errors in the line measurements compared to 
Paper I and the light curves of Panagia et al.  (1991) and Sanz 
Fern\'andez de C\'ordoba (1993).  The 
evolution of the line fluxes, leading to times for turn-on and maximum of the line 
emission, enable us to determine the absolute dimensions of the emitting gas 
independent of the imaging observations.  The implications of these results in 
connection with ground-based and \HST observations are discussed.  In \S 2 
we discuss the observations, in \S 3 the measurements of the line fluxes, 
and in \S 4 the main observational results are given.  The paper concludes 
with a discussion (\S 5) of ring geometry, abundances, density, emitting 
mass, and comparison with existing ring models in light of the 
observational data. 

\section{OBSERVATIONS}

The ultraviolet spectra presented in this paper were obtained with the 
\IUE SWP camera between 1987 March 14 and 1992 April 20.  All spectra 
were taken in the $10\arcsec\times 20\arcsec$ short wavelength large 
aperture (SWLA), the minor axis of which is approximately parallel 
with the spectrograph's dispersion direction.  The unique 
characteristics of the \sn\ field demanded special care during the 
acquisition process to ensure accurate target centering in the SWLA, 
as discussed below.  The observations were carried out at both \IUE 
Observatories, located at NASA/Goddard Space Flight Center and ESA's 
Villafranca ground station (VILSPA).  The SWP image number, time of 
observation, and exposure duration for each low dispersion spectrum 
are listed in Table 1.  The observation date refers to the starting 
time of the exposure, expressed in days after outburst.  The outburst 
time is assumed to be that of the IMB neutrino detection (1987 
February 23.316 UT, Bionta et al.  1987).

Twenty SWP high dispersion spectra of \sn\ were obtained during 
the UV emission line phase $(75<t<1800)$ days.  These observations and 
their analysis are discussed below in \S 4.4.

Following the rapid decrease of the supernova's UV flux shortward of 
2000 \AA\ in the first week of the outburst, the continua of two 
early-type stars appeared in SWLA spectra of \sn\ (Kirshner et al.  
1987).  The identification of the progenitor hinged on close scrutiny 
of these two previously unknown stars in the LMC located within 
arc seconds of the supernova's position.  Analysis of their 
spatial separation in SWP low-dispersion spectra taken several weeks 
after outburst showed that Sk $-69\arcdeg\,202$ had disappeared 
(Gilmozzi et al.  1987; Sonneborn, Altner, \& Kirshner 1987).  These 
same two stars, referred to as star 2 and star 3, after their position 
in the tabulation of West et al.  (1987), have dominated UV flux in 
SWP large aperture exposures of \sn\ since 1987 March.  Stars 2 
and 3 are offset from \sn\ by 2\farcs91 (p.a.  318\arcdeg) and 
1\farcs63 (p.a.  118\arcdeg), respectively, as measured in recent 
\HST WFPC2 images (Romaniello, 1996, private communication.  These 
updated positions are within the measurement errors of 
Walborn et al. (1987).

Several known sources contribute to the observed spectrum in the \IUE 
large aperture exposures taken at the position of \sn.  In 
addition to the supernova, there is the emission from circumstellar 
material (the subject of this paper), the continua of stars 2 and 3, 
and a very faint star located about 9\arcsec\  from \sn\ which 
enters the end of the SWLA every six months when the aperture has the 
appropriate orientation.  These contributions were taken into account 
in the measurement of emission line strengths.  Stars 2 and 3 are the 
most important secondary sources for purposes of this paper.  

The shape of emission lines and supernova continuum are affected by the 
detailed structure of the combined stellar spectra.  Correcting the 
observed SWP spectrum requires more than the subtraction of a stellar 
background spectrum.  By virtue of its location near the south 
ecliptic pole, the LMC is observable  365 days per year and the 
position angle of the \IUE large aperture changes by about one degree per 
day.  As the orientation of the SWLA rotates on the sky during the 
year, the projection of stars 2 and 3 on to the spectrograph 
dispersion axis also varies.  Stars 2 and 3 appear to rotate about the 
supernova's position (relative to the aperture) and their projection 
onto the direction of spectral dispersion has a 365-day periodicity.  
The spatial separation (4\farcs5) corresponds to $\sim 1200 \kms$ 
($\sim$7 \AA) when the stars are aligned parallel to the dispersion 
direction.  The principal spectral features in the stellar spectra 
shift in wavelength because star 2 contributes about 70\% of their 
combined flux below 2000 \AA.

Accurate centering of the supernova in the SWLA was necessary in order 
to obtain reliable emission line light curves.  During the time 
interval spanning the observations, \sn\ covered over 13 magnitudes of 
visual brightness.  The \IUE Fine Error Sensor (FES) was used to 
acquire the supernova, providing position measurements of its center 
of light from day 1 (V$\sim 4.5$), through visual maximum (day 86, 
V$\sim3.0$), and up to day $\sim 885$ (V$\sim13.4$).  After that time 
the supernova was acquired as a blind offset from the nearby star Sk 
$-69^{\circ}\,203$, 138\farcs5 almost due N of \sn\ (p.a.=2\fdg63).  
The coordinates (B1950.0) used for the blind offset maneuver are 
$\alpha=05^{\rm h} 35^{\rm m}$ 48\fs8, $\delta =-69\arcdeg 15\arcmin 
39\farcs4$ for Sk $-69\arcdeg\,203$ and $\alpha =05^{\rm h} 35^{\rm 
m}$ 50\fs0, $\delta =-69\arcdeg 17\arcmin 58\farcs0$ for \sn.  The 
offset maneuver was tested and calibrated between days 700 and 850 
when \sn\ was still bright enough ($12 \lesssim V \lesssim 13$) to 
directly acquire it with the FES.  The measured maneuver errors on 
these occasions were $\pm 1$ FES unit in each axis, or $\pm$0\farcs3 
total pointing error on the plane of the sky.  We therefore believe 
that there were negligible differences in aperture centering over the 
five years of the observations which might affect the emission line 
spectrophotometry.

\section{DATA ANALYSIS}

The procedures used to measure the emission line strengths in low 
dispersion spectra are described in this Section.  All line fluxes 
were measured relative to the net supernova spectrum, that is the 
observed spectrum with the contribution of stars 2 and 3 subtracted.  
This procedure, which in principle is the same as used in Paper I, 
assumes that the relative contributions of stars 2 and 3 to the 
observed spectrum are constant.  This will be true in general provided 
that the supernova is centered in the SWLA so that both stars are also 
in the aperture and that the stars themselves are not highly variable in the 
ultraviolet. This last point is addressed in \S 4.1.
The principal steps in this process are outlined below.

\subsection{SPATIALLY RESOLVED SPECTRAL DATA}

Our analysis uses the spatially resolved spectra from the  extended
line-by-line (ELBL) data files produced by the standard IUESIPS processing
(Turnrose \& Thompson 1984, as updated by Munoz Piero 1985 for ELBL
processing). The ELBL file is a two-dimensional spectral image 
extracted from the \IUE photometrically corrected image along lines
of constant wavelength. For the SWP camera it contains about 700 points in the
wavelength domain (1150--1980\,\AA,  $\sim 1.2\,$\AA/pixel) and 110 points in
the spatial dimension (0\farcs68 per linear pixel).  

Specialized interactive software was written to correct ELBL pixels 
corrupted by energetic particle hits and bright spots.  For each image the 
area inside the extraction slit and in the background areas were inspected 
for bright spots and cosmic ray effects.  When found in the background 
extraction slit, corrupted pixels were replaced with with data values 
corresponding to the average of immediately adjacent pixels.  Extreme care 
was exercised when correcting bright spots close to the spectrum.  Within 
the extraction slit bright spots were replaced with the average of part of 
a single row of pixels (parallel to the dispersion) at the same distance 
above or below the dispersion line.

The spectrum was extracted from the corrected image  using a
rectangular extraction slit, similar to the standard IUESIPS extraction
procedure described by Turnrose \& Thompson (1984).  
The spatial height of the rectangular spectral extraction slit for the SN 
1987A spectrum was 24 lines (16\arcsec).  For comparison, the IUESIPS 
standard slit heights are 18 and 30 lines for point source and extended 
source extractions, respectively.  The extraction slit height of 24 lines 
was chosen to include the wings of the point spread function of stars 2 and 
3 at their maximum spatial separation in the spectrograph format, i.e., 
when their position angle was perpendicular to the dispersion direction.  
The customized slit height minimizes the background contribution to the 
extracted spectrum.  The standard point source extraction slit is too short 
while the standard extended source extraction slit is longer than 
necessary.  Use of a slit-weighted extraction technique, such as 'optimal' 
(Kinney et al.  1991) was not used because of the variable, 
non-point-source nature of these spectra.

The extracted spectra were corrected for two calibration changes. First, a
secular shift in the wavelength scale between 1984 and 1988, and second the
long-term decline in SWP sensitivity. Starting in 1984 a systematic error
developed in the SWP low dispersion wavelength scale and grew linearly with
time.  The error resulted from a slow divergence of true dispersion 
relation from the extrapolation of the mean time-dependent dispersion 
constants.  New dispersion constants were implemented in IUESIPS in 1988 
April at GSFC and 1988 September at Vilspa (Thompson 1988).  The SWP low 
dispersion spectra processed between 1987 February and 1988 September 
contained an error in the wavelength calibration amounting to $\sim 
2.5$~\AA.  Thompson (1988) provided an algorithm to fit the time-dependent 
error in the wavelength scale.  This relation was used to correct all SWP 
low dispersion spectra of \sn\ effected by this calibration change.

Long-term changes in the sensitivity of the \IUE instrument were 
compensated for by using the correction scheme of Garhart (1992a).  
Between 1225 and 2000~\AA\ the SWP low dispersion sensitivity has 
decreased linearly since 1979.5 at wavelength-dependent rates of 
$-0.2$\% to $-1.2$\% per year.  The Garhart sensitivity correction 
method gives results similar to Bohlin \& Grillmair (1986) for the 
period in which they overlap.  The extracted fluxes were also 
corrected for the temperature dependence of the SWP camera response 
($-0.5$\% per \arcdeg C, Garhart 1992b).  Finally, the spectral fluxes 
were placed on the \IUE absolute flux scale using the 1980 absolute 
calibration (Bohlin \& Holm 1980).

\subsection{REFERENCE SPECTRUM OF STARS 2 AND 3}

There was a period of a few weeks in 1987 when the SWP large aperture 
spectrum of \sn\ was due almost entirely to stars 2 and 3.  This 
occurred between mid-March and early April, before the emission lines 
appeared.  The SWP large aperture spectra taken during this period 
were examined to determine which ones could be combined to form a 
reference spectrum of stars 2 and 3 .  Criteria included no 
overexposure, good signal-to-noise at wavelengths shortward of 
1500~\AA\ (i.e., $t_{exp}\gtrsim 160$min), and no apparent signal from 
\sn, particularly longward of 1800~\AA.  Three spectra were 
selected: SWP 30592 (day 27.3, $t_{exp}=300$ min), SWP 30637 (day 
33.1, $t_{exp}=185$ min), and SWP 30743 (day 44.2, $t_{exp}=225$ min).  
These spectra were reduced in the same manner described in \S3.1.  SWP 
30637 and 30743 were registered in wavelength with respect to SWP 
30592 by cross correlation.  The mean spectrum is the average of the 
three weighted by the individual exposure times.  The resulting SWP 
spectrum of stars 2 and 3 is shown in Figure 1.  This spectrum is 
similar, though not identical, to that used by Pun et al.  (1995) for 
their study the UV continuum properties of the \sn\ outburst.  They 
derived their star 2 and 3 UV spectrum by averaging {\it late}-time 
\IUE spectra ($1300<t<1500$ days), when the emission lines were 
extremely weak, in order to recover the UV spectrum of the supernova 
early in the outburst.  In contrast, we use {\it early}-time spectra 
for stars 2 and 3, before the emission lines appear, so that we can 
follow the decline of the emission lines at late times.  Our spectrum 
of stars 2 and 3 is in good agreement with the FOS spectrum of star 2 
(Fig 1 of Scuderi et al. 1996), given the differences in spectral 
resolution and calibration between the two instruments.  Star 2 
contributes $\sim$70\% of the flux in this wavelength region.

\subsection{BACKGROUND STAR SUBTRACTION} 

Subtraction of the reference spectrum of stars 2 and 3 from the 
observed spectrum requires that the two wavelength scales be 
registered to correct for any offset.  Such a wavelength shift could 
be due to either an offset in the center of light from the center of 
the SWLA or a shift in the spectral features of stars 2 and 3 in the 
observed spectrum.  In the case of the former, a 1\farcs0 pointing 
error in the dispersion direction corresponds to a spectral shift of 
$\sim 300 \kms$, or $\sim 1.6$~\AA.  The wavelength shift required to 
properly align stars 2 and 3 with the observed spectrum was computed 
with an interactive cross-correlation program from the NASA/GSFC \IUE 
Data Analysis Center software library.

Figure 2 shows two SWP spectra with aperture position angles differing 
by 156\arcdeg.  The shifts in the observed spectra, relative to the 
reference spectrum of stars 2 and 3, are evident, as is the difference 
in the widths in the strong stellar lines.  Relative to the reference 
spectrum, the spectra are shifted by $-459$ and $+243 \kms$, as 
determined by the cross-correlation analysis.  Figure 3 shows the 
wavelength registration shifts for the \sn SWP data set (Table 1).  
There is a $\sim 1$ year periodicity in the data, indicating that the 
position of stars 2 and 3 relative to the spectrograph-aperture 
geometry is a major source of this variation.  Within about 
$\pm$0\farcs5 ($\sim \pm 150\kms$) the wavelength registration shifts 
are dominated by target acquisition and pointing errors.  The net 
spectrum is computed by subtracting stars 2 and 3 from each observed 
spectrum after shifting the spectrum of stars 2 and 3 by the amount 
determined by the cross correlation and interpolating it to the 
wavelength scale of the observed spectrum.

\section{EMISSION LINE RESULTS}

\subsection{FLUXES AND LIGHT CURVES}

The  flux in each emission line was measured in the net spectrum 
where local continuum determination is simplified for most of the 
emission lines of interest.  The subtraction of stars 2 and 3 removes 
strong stellar and interstellar absorption features, including 
\SiII\wl1260, \SiIII\wl1300, \CII\wl1335, and \AlIII\wl1670. 
The wavelengths defining the center and red and blue edges of each 
line in each spectrum were determined interactively.  After 
subtracting stars 2 and 3 the local continuum was assumed to be flat, 
except where the SN UV flux is strong.  The latter is confined to $\lambda
\gtrsim 1700$~\AA\ and $t\lesssim 700$ days.  The line strength was 
integrated by a trapezoid algorithm.  An example is 
shown in Figure 4 where the observed and net spectra near the \NV\ line 
in SWP 36801 (day 894) are 
shown on the same flux scale.  The \NV\ line is easily distinguished in 
the net spectrum.

Table 1 contains the fluxes for the \NV\wl1240, \NIVb\wl1486, 
\HeII\wl1640, \OIIIb\wl1665, \NIIIb\wl1750, and \CIIIb\wl1908 lines 
measured in 119 spectra taken on 106 dates.  The $1225-1980$~\AA\ 
history of \sn\ between days 85 and 1820 is shown in Figure 5 where 
the observed net fluxes have been averaged into nine time intervals.  
The development of the six emission lines is clearly seen.  In 
addition to these emission lines, other features are evident in the 
net spectra.  The supernova continuum longward of 1700~\AA\ is 
prominent until day $\sim700$ (see also Pun et al.  1995).  There are 
additional features which rise and fall in the same time period 
between 1280 and 1540~\AA.  Of particular interest is the narrow 
feature at 1400~\AA, which we tentatively identify with 
\OIVb\wll1397-1407.  Although we have no identifications for the other 
features in the spectra, they are believed to be part of the \FeII 
blanketed continuum of the supernova.  The formation of such 
"pseudo-emission" features in the early spectra of \sn\ is discussed 
by Cassatella et al.  (1987) and Lucy (1987) and at late times by 
Fransson (1994) and Li \& McCray (1996).  Local maxima occur where 
line blocking is less important and the photons can escape, i.e., at 
wavelengths where there are gaps in the list of strong lines.  Similar 
pseudo-emission features occur in classical novae.  Hauschildt et al.  
(1994) modelled and discussed the formation and evolution of these 
features in Nova Cygni 1992.

The resulting light curves for the \NV, \NIVb, \HeII,
\OIIIb, \NIIIb, and \CIIIb\  lines are shown in Figure 6. 
The light curves show the observed line fluxes, without correction for 
interstellar extinction (see \S 5.2).  It was sometimes difficult to 
determine the continuum level at the base of the \CIIIb\ emission line, even 
after subtracting stars 2 and 3, because of the UV flux from the supernova 
debris increased in this region of the spectrum (see Fig.  5).  A check of 
the accuracy of the low dispersion \CIIIb\ flux measurements was made by 
comparing them with \CIIIb\ fluxes from high resolution observations where 
continuum determination relative to the narrow emission is much easier.  
The high dispersion data for \sn\ are discussed below (\S 4.4) and the 
corresponding \CIIIb\ fluxes also shown in Figure 6d.  The 
good agreement between these independent measurements, well within the 
scatter of the low resolution data, illustrates the overall accuracy of the 
continuum determination.  The new emission line light curves are 
clearly superior to those published previously (Paper I, 
Sonneborn et al. 1990, Panagia et al. 1991, and Sanz Fern\'andez de 
C\'ordoba (1993) and may effect results derived from them, for example Gould 
(1994, 1995).

The size of the \IUE large aperture prompts us to consider the degree 
to which emission from the outer rings might contribute to the UV 
light curves, particularly at early times (say, $t<600$ days).  If one 
adopts the geometry suggested by Burrows et al.  (1995) the rising 
branch of the light curve of the southern outer ring (SOR) should 
extend from about one month after explosion to about three years 
later, whereas the northern outer ring (NOR) should have become 
visible about two years after outburst.  In this geometry the NOR 
would not have been entirely visible in the earliest images (Crotts, 
Kunkel, \& McCarthy 1989) but may have been "complete" by late 1989 in 
the ESO images (Wampler et al.  1990).  The UV light curves do show a 
weak secondary maximum around 1000 days, consistent with the epoch 
when both outer rings would have reached their maximum emission 
(Panagia et al. 1996, in preparation).  Therefore, emission from the 
NOR would not contribute to the early \IUE observations, and emission 
from the SOR would have contributed less than 1/3 of its total 
strength.  Present day \HST imaging shows that the NOR, the SOR and 
the inner ring have comparable total emission in optical lines.  
However, the average density in the inner ring is higher than that in 
the outer rings so the time decay of the inner ring emission would be 
faster than that of the outer rings.  As a consequence, the total 
emission of the inner ring at the time of its maximum must have been 
considerably higher than that of the outer rings combined.  In fact, 
ESO observations around the time of the tentative secondary maximum 
show that the outer rings contribute less than 10\% of the total 
[\OIIIb emission (J.  Wampler \& L.  Wang 1992, private 
communication).  We therefore believe that the contribution of the 
outer rings to the \IUE emission line light curves is negligible.

Figure 4 shows that the subtraction of stars 2 and 3 is not perfect.  
Due to fixed-pattern and other small-scale noise, the scatter in the 
net spectrum near the emission line limits the accuracy of continuum 
subtraction, and hence the significance of short-term fluctuations in 
the emission line light curves.  The accuracy of the continuum 
subtraction is quantified by the parameter $\delta F_c$, the local 
continuum residual (Eqn.  1), which is the difference between the net 
spectrum and the adopted local continuum, averaged over a bandpass 
comparable to the width of the emission line,
\begin{equation}
\delta F_c = {1\over{\lambda_2 - \lambda_1}}\int_{\lambda_1}^{\lambda_2} 
\left(F_{net}(\lambda) - F_{local}(\lambda)\right) ~d\lambda.
\end{equation}
$\delta F_c$ was measured near the \NV\ line because the adjacent 
continua of stars 2 and 3 have strong stellar and interstellar 
features, which if not subtracted accurately will add significant and 
erroneous structure to the net spectrum.  Perfect continuum 
subtraction would yield $\delta F_c=0$.  $\delta F_c$ was computed in 
a bandpass centered at 1262~\AA\ with a full width of 10~\AA, the 
characteristic width of the \NV\ emission line at zero intensity.  
Figure 7 shows the \NV\ light curve (same data as Fig.  6a) and 
$\delta F_c$.  The error bar $\pm\sigma=1.82 \EE{-14}\,\ergcms$ 
corresponds to the measurement uncertainty of individual data points 
for the six emission lines, indicating that the continuum 
subtraction procedure works well over the entire 5-year span of the 
data set. 

The accuracy of the continuum subtraction is insensitive to the known 
variability of the two stars.  First, a change in the continuum flux 
results in a vertical shift of the net spectrum.  The continuum in the 
net spectrum is constant (i.e. horizontal in Fig. 4), 
except in the presence of the strong SN continuum, so that the 
emission line is measured relative to this shifted continuum level.  
Second, it is now well established that only star 3, a Be star, shows 
any significant variability.  Walborn et al.  (1993) showed that star 
3 has photometric variability ranging from $\pm0.15$ mag in $U$ to 
$\pm0.3$ mag in $R$ and $I$.  Since star 3 contributes only $\sim30$\% 
of the UV flux, this level of variability is within the measured value 
of $\sigma$.

The measurement uncertainty of the line fluxes implies that the 
significant short-term fluctuations in the light curves are 
astrophysical in origin, and are not instrumental effects.  The increased 
scatter in the \NV\ and \NIIIb\ light curves between day 900 and 1100, 
in the \NIVb\ line before day 400, and the large values of \NIIIb\ and 
\NV\ near maximum are actual fluctuations in the emission line fluxes.  
We find no evidence that these spectra are affected by cosmic ray hits 
or other spurious effects and conclude that the fluctuations are real.  
The largest \NV\ flux point (day 485) remains suspect, however, as it 
occurs significantly after maximum brightness and there is no related 
response in the other lines.

\subsection{EMISSION LINE MAXIMUM}

The times when the lines first appear and when they reach maximum 
brightness are two important light curve parameters used to estimate 
the size and inclination of the ring.  With a circular ring geometry 
the emission lines are expected to appear suddenly as the light echo 
paraboloid makes first contact with the ring (Dwek \& Felten 1992).  
In general, the highest ionization stage available should be used to 
measure these parameters.  

It is evident from Figure 6 that the six UV lines reach maximum at 
approximately the same time.  However, only the \NV\ and \NIIIb\ lines 
have large fluxes and well-defined maxima necessary for a more 
quantitative analysis.  The actual times of maxima are $t^{NIII}_{max} 
= 399 \pm 15$ and $t^{NV}_{max} = 429\pm 19$ days ($418\pm 12$ days if 
the high data point at day 485 is excluded), as determined by 
high-order polynomial fits to the light curves.  Comparison of the 
\NV\ and \NIIIb\ in Figure 8 shows that while their post-maximum 
decline is very similar, the light curves before maximum have 
significant differences.  Figure 8 shows a large deficiency of \NV\ 
relative to \NIIIb\ between 150 and 400 days and that \NV\ peaks about 
three weeks after \NIIIb.

The \NV\wl1240 emission 
from the far side of the ring may be delayed by 1--2 weeks by 
resonance scattering in the ring and the bipolar nebula, as discussed 
by LF96.  However, to obtain an even longer delay, and thus later 
$t^{NV}_{max}$, like the one observed, additional resonance scattering is 
needed.  The most likely source is in the \HII region interior to ring 
discussed by Chevalier \& Dwarkadas (1995).  Preliminary calculations 
by us show that there may be a sufficient optical depth in \NV\wl1240 
through this region along the line of sight to the far side of the 
ring. The large deficiency of \NV\ relative to \NIIIb\ is therefore probably due 
to resonant scattering of \NV\ photons by N$^{+4}$ ions in and outside 
the ring.  This biases the \NV\ flux maximum and complicates its 
interpretation by introducing effects other than light travel time and 
local recombination.  In view of the uncertainties 
in interpreting the \NV\ light curve, we believe 
$t^{NIII}_{max}$ provides the best measure of the time of maximum, 
$t_{max}$, and we use this in our subsequent analysis.

\subsection{EMISSION LINE TURN-ON}

Figure 9 shows the \NIIIb, \NIVb, and \NV\ fluxes up to day 120.  All 
three lines appear very abruptly, as predicted.  There is also an 
ionization dependence in the turn-on times.  \NV\ appeared first, 
after day 70.2 and before day 80.6.  \NIVb\ appeared next, between 
80.6 and 85.2 days.  \NIIIb\ appeared last, between 85.2 and 90.9 
days.  While it is possible that \NIIIb\ was present at a very weak 
level between day 70 and 85, the measurements are consistent with zero 
flux based on the scatter before day 60.  Since the semi-forbidden 
lines are not effected by scattering, \NIVb, the next highest 
ionization state after \NV, should closely track the inside edge of 
the ring.  However, as discussed below (see also Fig. 6 in LF96), 
the \NV\ light curve is significantly affected by resonance scattering   
which introduce large systematic errors. Consequently we believe 
\NV\ is also inappropriate for measuring $t_{rise}$. 
We take the sharp rise in the \NIVb\ flux as the turn-on 
time and find $t_{rise}=83 \pm 4$ days (estimated error).  The 
uncertainty in the turn-on time is set by the spacing of the 
observations, and consequently $t_{rise}$ may not be resolved in the 
\IUE data.

On the near side of the ring resonance scattering of \NV\ by N$^{+4}$ 
ions in the ring and any other intervening highly ionized 
circumstellar gas should delay the turn on of this line compared to 
that of other nitrogen lines.  Instead, the observations show an 
earlier turn on of \NV.  The most likely explanation to this apparent 
inconsistency is the ionization stratification of the ionized gas.  
The effect of stratification on the turn on of \NV\ should be larger 
if there is an \HII region interior to the ring from which there also 
may be weak \NV\ emission.  Another possible source of early \NV\ 
emission is the bipolar nebula close to the ring, but away from its 
equatorial plane where the nebula may not be ionization bounded.  
This part of the nebula may have a lower density 
than the ring causing it to glow early in \NV, but not in other UV 
lines due to long recombination time scales.

\subsection{HIGH DISPERSION SPECTRA}

The first SWP high dispersion spectrum of \sn\ during the narrow 
emission line phase was taken on 1987 Nov.  25 (day 275) and used in 
Paper I.  Nineteen additional high dispersion SWP spectra were 
obtained between then and day 1652.  The observations are summarized 
in Table 2.  Using standard IUESIPS high dispersion processing, 
extracted spectra were calibrated using the flux calibration of 
Cassatella et al. (1994), which provides an accuracy of about 4\% 
with respect to low resolution spectra longward of $\sim1400$~\AA\ and 
increasing by no more than a factor of two at shorter wavelengths.  
All the lines seen in low resolution were also detected in high 
dispersion spectra taken near peak brightness.  However, only the 
\NIIIb\ and \CIIIb\ lines were detected over a sufficiently long time 
base and with sufficient signal-to-noise to be able to discuss their 
evolution and compare them with the low dispersion data.

Table 2 also contains the fluxes for \CIIIb\wll1906.68, 1908.73 
and \NIIIb\wll1749.68, 1752.16.  The other members of the 
\NIIIb\ multiplet were either too weak or, in the case of \wl1748.61, 
badly effected by a camera reseau. Figures 10 and 11 show the spectra
for these lines. The line widths are consistent with instrumental 
broadening (FWHM $\sim0.18$~\AA\ at \NIIIb\ and $\sim$ 0.20~\AA\ at 
\CIIIb) for the time period where line widths are reliably measured 
(up to day $\sim$1000 for \CIIIb).

The relative strengths of the \CIIIb\ and \NIIIb\ multiplet components 
are well known diagnostics of electron density (Nussbaumer \& Schild 
1979; Nussbaumer \& Storey 1979; Berrington 1985).  In Paper I we 
showed that $n_e = 1-3 \times 10^4 \cm3$ at the time of the first high 
dispersion spectrum in Table 2 (day 275).  The larger number of 
spectra now available makes it possible to examine the line ratios for 
evidence of density variations.  Figure 12 shows the \CIIIb\  
\wl1907/\wl1909 and \NIIIb\  \wl1750/\wl1752 line ratios before day 
1000.  Calculated $I_{1907}/I_{1909}$ intensity ratio as a function 
of electron density are shown in Figure 13 for three temperatures 
using a 6-level \CIII\  atom and recent atomic data.  The \CIIIb\ line 
ratio of $0.8 \pm 0.2$ up to day 408, implies $n_e = 4.7\pm2.0 \times 
10^4 \cm3$, assuming $T_e \sim 50,000\rm\,K$. After $\sim 
600$ days, $I_{1907}/I_{1909}\sim 1.5\pm0.4$.    
The average electron density for the \CIIIb\ emitting gas decreases by 
a factor of $\sim3$ or more from $\sim 5 \times 10^4 \cm3$ before 
$\sim400$ days to $\lesssim 1-2 \times 10^4 \cm3$ thereafter, 
depending on $T_e$. Near $\sim 600$ days $T_e \sim 30,000\rm\,K$ may 
be more appropriate, favoring the lower estimate of $n_e$.  The 
increasing uncertainty in the later data points makes the line ratios 
after $\sim800$ days unreliable for density analysis.
 
\section{DISCUSSION} 
 
The absolute dimensions of the emitting region can be derived 
from the shape of the UV light curves with some basic assumptions 
about its geometry, mainly a circular thin ring. For this purpose we summarize 
the main features of the recombination/cooling -- light echo scenario.
Hydrodynamic models of the breakout of the supernova
shock wave through the photosphere of the progenitor show that a strong pulse
of soft X-rays is emitted with a duration  of a few hours (Shigeyama,
Nomoto, \& Hashimoto 1988; Woosley 1988; Shigeyama \& Nomoto 1990;
Blinnikov \& Nadyozhin 1991; Ensman \& Burrows 1992).
The photospheric temperature, and also the radiation temperature, 
decrease from  $\sim 10^6 \rm\, K$ to $\sim 2\EE{4} \rm\, K$
in about 12 hours. The ionizing flux from the supernova is therefore 
exclusively emitted during the first day, and in contrast to the 
interstellar visual light echoes, we can consider the emission from 
the supernova as instantaneous. The soft X-rays ionize the
circumstellar gas on a time scale of hours, leading to highly ionized 
species like the observed \NV, or even higher (LF91; LF96). 
At the same time the radiation heats the gas to temperatures 
exceeding $\sim 10^5 \rm\, K$. The subsequent evolution of the gas, 
occurring on a time scale of weeks to years, is determined by the 
coupled cooling and recombination of the plasma. 

\subsection{GEOMETRY}

The recombination and cooling times are set by the temperature and 
density, $t_{rec} \sim 1 / (\alpha_{rec} n_e)$, so gas at low density 
recombines and cools on a longer time scale.  The finite velocity of 
light implies that only a region bounded by the light echo paraboloid, 
given by $r =c t /(1 - \cos {\theta})$, can be observed from the Earth 
at time $t$.  Here $r$ is the distance from the supernova, $t$ the 
time of the observation, and ${\theta}$ the angle between the line of 
sight and the direction to the emitting gas as seen from the 
supernova.  In the case of a spherical shell, with radius $R_s$, the 
whole structure is seen after a time $t_{ max}= 2 R_s/c$, while for a 
circular ring $t_{ max}= (1 + \sin i) R/c$, where $R$ is the ring's 
radius and $i$ is the inclination of the plane of the ring with 
respect to the line of sight ($i=0\arcdeg$ is face-on).  For ions 
which recombine and cool slowly, the observed flux increases nearly 
linearly up to $t_{max}$.  If the density is high enough recombination 
occurs rapidly and the light curves of different ions reach maximum 
simultaneously followed by decay on the individual recombination time 
scales.  In the slow recombination case the 
light curves will remain at a roughly constant level after maximum.

LF91 showed that for a {\it spherical} shell the gas was marginally 
optically thin in the continuum.  In the case of a {\it ring} 
geometry, however, the gas must have significant optical depth in the 
EUV continuum because of the larger column density needed to 
produce the same emission (LF96).  If the ionization of the gas were 
radially constant (as is probably {\it not} the case), the light 
curves of medium 
ionization ions formed by recombination, like \NIII, would be delayed 
with respect to those formed by photoionization at the time of the 
outburst, like \NV.  However, from the outset the ring will be optically thick 
to ionizing photons, resulting in a radial ionization gradient with several 
ionization zones (Lundqvist 1992).  Medium ionization ions like \NIII\ and 
\NIV\ are created with high abundances outside the \NV\ -- \NVI\ zone.  
Models using a ring geometry are discussed in detail by LF96, and are 
consistent with the observed light curves.  The delayed turn-on of
\NIVb\ and \NIIIb\ (Fig.  8) relative to \NV\ supports this model.
  
The \HST images clearly resolve the ring-like shape of the emitting 
gas in line and continuum radiation (Jakobsen et al.  1991, 1994; 
Plait et al.  1995; Burrows et al.  1995).  While the extent of the 
region perpendicular to the ring plane remains to be determined by 
modelling of the lines, there are additional indications from the \IUE 
observations for a thin ring geometry.  The most direct evidence for 
this comes from the abrupt rise of the emission line fluxes.  As 
discussed in \S 4.1, we adopt the \NIVb\ turn-on time 
($t_{rise}=83\pm4$ days) as the initial rise time.  In \S 4.1 we also 
determined $t_{max}= 399\pm15$ days from the \NIIIb\ line.  The delay 
from an inclined circular ring is given by $t_{rise} = (1 - \sin i) 
R/c $.  We therefore have $R = c (t_{max}+t_{rise}) / 2= (6.24\pm 
0.20) \EE{17}$ cm, and $\sin i = c (t_{max}-t_{rise}) / 2 R = 0.656 
\pm 0.039$, or $i = 41\fdg0 \pm 3\fdg9$.  The inclination has also 
been measured from the \HST FOC images, assuming a circular ring.  
Jakobsen et al.  (1991) found $i = 43\arcdeg\pm 3\arcdeg$ and Plait et 
al.  (1995) found $i = 44.0\arcdeg\pm 1.0\arcdeg$.  The agreement 
between these values shows that the ring is nearly circular (see also 
Gould 1994; Lundqvist 1994; Crotts, Kunkel, \& Heathcote, 1995).

Our value of $R$ has direct bearing on the estimated distance to 
\sn\ using the method devised by Panagia et al.  (1991), and 
refined by Gould (1994, 1995).  Gould finds $t_{rise} = 75.0 \pm 2.6$ 
days and $t_{max} = 390 \pm 1.8$ days, which combined with the angular 
size of the ring in [\OIIIb\  at $t > 1200$ days from Plait et al.  (1995) 
gives a distance, $d_{SN} < 46.77 \pm 0.76\rm\, kpc$.  Our slightly larger 
values of $t_{rise}$ and $t_{max}$ and more realistic errors translate directly into a 
larger value and uncertainty for the distance, $d_{SN} = 48.6 \pm 
2.2\rm\, kpc$.  A more complete statistical treatment of the 
light curves and the time of maximum and related systematic 
errors effecting the distance to the supernova is beyond the scope of this paper 
but is discussed by Lundqvist et al. (1996, in preparation). 
We emphasize that systematic errors may be 
significantly larger than statistical errors for several reasons.  
For example, the method assumes perfect 
circularity of the ring.  Although this may be a reasonable 
approximation for the overall geometry of the ring, there may be local 
variations of the radius around the ring.  In particular, a small 
inward kink or near-absence of gas on the far side of the ring would 
give a smaller $t_{max}$ than for a circular ring, thus 
underestimating $d_{SN}$.  Second, the structure of the [\OIIIb\  ring 
at $t > 1200$ days was probably different from the structure of the 
gas emitting the UV lines at $t \lesssim 400$ days.  Indeed, LF96 
found that the ionization structure of the ring changes dramatically 
in only a few hundred days.  The UV lines are somewhat more sensitive 
to temperature and somewhat more peaked towards the inner edge of the 
ring than forbidden line emission (e.g., [\OIIIb).  The analyses of 
Panagia et al.  and Gould may be effected by a systematic error 
arising from the assumption that the optical forbidden lines and UV 
lines come from the same gas (but not observed at the same epoch).  
Depending on the radial distribution of the density in the ring, the 
[\OIIIb\  structure could either have a smaller, or larger, mean radius 
than the early UV line emission.  We also point out that their analyses 
used UV line 
fluxes from two independent reductions (labeled {\it GSFC} and {\it 
Vilspa} shown by Panagia et al.  (1991) of the {\it same} set of 
spectra, thus the number of independent data points is too large by a 
factor of two.

\subsection{ABUNDANCES}

The increased number of observations compared to Paper I, reduced 
measurement and systematic errors, and improved signal-to-noise ratios make 
it important to repeat the abundance analysis.  In order to convert the 
ionic ratios obtained from the line ratios to elemental abundances, 
assumptions about these relations have to be made.  More specifically, in 
Paper I it was assumed that N/C = \NIII/\CIII\  and N/O = (\NIII\ + 
\NIV)/\OIII.  The calculations by LF91 and LF96 show that, because of a 
balance of recombinations and collisional ionizations, this is a good 
approximation until the decay of the lines begin.  After this epoch the gas 
temperature is too low for collisional ionization to be important for 
\CIII, \NIII, and \OIII, and the lines decay on their respective recombination time 
scales.  These are quite different for the ions involved, and consequently 
the relations above break down.  Before $\sim 200 $ days the lines were 
weak and the fluxes therefore uncertain.  Consequently, in this analysis we 
restrict ourselves to $200 < t < 500$ days.  The collision strengths are 
taken from Berrington (1985), Keenan et.  al.  (1986), Blum \& Pradhan 
(1992).

The extinction toward \sn\ has been discussed in several papers and there
is now general agreement.  Walker \& Suntzeff (1990) have measured the
reddening in the field of \sn\ by studying the $(U-B)$ and $(B-V)$ colors
of 23 early-type stars in the vicinity of the supernova.  They determine the
reddening to be $\EBV_{total} = 0.17 \pm 0.02.$  Fitzpatrick \& Walborn (1990)
obtained optical and UV spectra of the LMC B0.7 Ia supergiant 
$\sim$2\arcmin\ from \sn\ (cf. \S 2).  They conclude that 
$\EBV_{total} = 0.16$ for \sn, with Galactic and LMC components 
of $\EBV_{Gal}=0.06$ ($A_{\small \lambda}$  from Savage \& Mathis, 
1979) and $\EBV_{LMC}=0.10$ ($A_{\small \lambda}$ for 30 Dor from 
Fitzpatrick, 1985). We adopt the Fitzpatrick \& Walborn extinction 
toward Sk $-69\arcdeg\,203$ for \sn. The total extinction correction
factors for the six emission lines are listed in Table 3. In Paper I 
we adopted $\EBV_{total} = 0.20$, the  extinction correction for which  
is also given in Table 3 ($\EBV_{Gal}=0.06$ and  
$\EBV_{LMC}=0.14$). The line luminosities are computed by correcting 
for extinction and the distance to the supernova. Assuming isotropic 
radiation and a distance of $50\rm\, kpc$, the dilution factor is $3.0\EE{47}$.

The ratios depend weakly on the temperature, and in Table 4 we give 
the resulting values of N/C and N/O as function of this parameter.  In 
Paper I we argued that $T_e \sim 5\EE4 \rm\,K$ from 
[\OIIIb\ observations (Wampler \& Richichi 1989) and the models in 
LF91.  Using this temperature we find that N/C~$ = 6.1 \pm 1.1 $ and 
N/O~$ = 1.7 \pm 0.5$, where the errors are purely statistical.
Accepting an (unrealistically large) uncertainty in $T_e$ from $10^4 
\rm\,K$ to 
$10^5 \rm\,K$, corresponds to a range in N/C of $4.7 - 15.7$.  The statistical 
errors in the N/C ratio are therefore considerably less than the 
systematic.  Because the lower limit is N/C $ = 4.7$, and a more 
realistic value is N/C $ = 6.1$, the conclusions in Paper I on the 
large nitrogen enhancement are unaltered. 
For comparison, LF96 derive N/C~$=5.0 \pm 2.0$ and N/O~ $=1.1 \pm 0.4$ 
from their photoionization/recombination model.  The two methods are 
in good agreement.
  
 There is  a weak 
tendency for the derived \NIII/\CIII\ ratio to decrease with time.  For 
$400 < t < 800$ days we find\ \NIII/\CIII\  $= 5.3 \pm 1.5$ for $T_e = 
5\EE4 \rm\,K$.  The decreasing \NIII/\CIII\ ratio is consistent with the 
expectations from the models in LF91, and justifies limiting the 
abundance analysis to $t \lesssim 500$ days, when N/C = \NIII/\CIII\ is 
a reasonable approximation.

The nebular analysis has a number of limitations.  A 
nebular model gives a reliable estimate of the ionization structure 
under steady-state conditions or immediately after the gas has been 
ionized.  As the ionizing radiation vanishes, the ionization state 
evolves away from this structure due to different recombination time 
scales for the different elements (C, N, and O).  For example, at 
$t_{max}$ the part on the very far side of the ring can be roughly 
explained by a nebular structure, but the gas on the very near side 
has had $t_{max}-t_{rise}$ days to evolve away from such a structure.  
Because the observed emission at $t_{max}$ is an overlay of emission 
from different parts of the ring which have had different times to 
evolve off the steady-state structure, a pure nebular analysis is 
bound to have some systematic error.  In contrast, LF96 did not have 
to assume an ionization structure for their models since it and the 
temperature are calculated in a self-consistent manner.  This results in 
smaller systematic errors related to ionization equilibrium. Given 
these uncertainties, the CNO abundance ratios in this paper and in 
LF96 are in good agreement. In the near future spatially resolved 
spectroscopy of the inner ring with STIS on \HST will make possible this 
type of nebular analysis as a function of position around the ring.

\subsection{EMITTING MASS}

>From the \HeII\wl1640 line we can estimate the mass of the \HeII 
emitting gas, i.e., the mass of \HeIII\  zone.  In order to avoid having 
to compensate for light travel time effects, we evaluate the \HeII\  
flux near day 400 so that the full volume is observed.  Near $t_{max} 
\pm 100$ days the observed \HeII\ flux was $(9 \pm 2)\EE{-14}\ergcms$ 
(estimated uncertainty), and the luminosity 
$1.0\pm 0.2\EE{35}\ergs$, assuming $\EBV=0.16$.  The 
\HeII\ emissivity does not vary much with time (LF91 and Fig. 6e ), so 
$L_{HeII} \approx \int j_{eff}(T_e)~ n_e~ n_{HeIII} ~dV $, where 
$j_{eff}$ is the effective emissivity of the \wl1640 line.  In the 
\HeIII zone $n_{HeIII} \approx n_{He}$.  Furthermore, Wang (1991) finds 
that $n_{He}/n_{H} \approx 0.2$, and we find for the total mass of 
this zone
\begin{equation}
M({\rm He~ III}) \approx 1.4\EE{-2}
\left({n_e \over 2.6\EE{4} \,\cm3}\right)^{ -1} 
\left({T_e \over 5\EE{4} \rm K}\right)^{ 0.87} ~ \Msun.
\end{equation}
This mass estimate only refers to the \HeIII\  zone. As models
by LF96 show, the total mass of ionized gas can be several times
larger. A better estimate of the emitting mass 
is obtained from the $\Hb$ luminosity.
Wampler \& Richichi (1989) estimate $L(\Hb) \approx 1.4\EE{34}\ergs$
on day 310 ($\EBV = 0.16$). Multiplying this by a factor 3/2, since
only $\sim 2/3$ of the total ring was within the light echo paraboloid at 
that time, and again using $n_{He}/n_{H} \approx 0.2$, we obtain a total 
mass of the \HII\  region
\begin{equation}
M({\rm H~ II}) \approx 4.7\EE{-2}
\left({n_e \over 2.6\EE{4}\,\cm3}\right)^{ -1} 
\left({T_e \over 5\EE{4} \rm K}\right)^{ 1.02} ~ \Msun
\end{equation}
on day 395. Because of the high temperature, both the
\HeII\ line and $\Hb$ are affected by collisional excitation. The
mass estimates are therefore upper limits to the actual masses.
(See LF96 for an estimate which takes these model dependent effects
into account.) Our estimates, however, suggest that the \HeIII\  zone
occupies $\sim 30\%$ of the total ionized region.

The first \HST observations of \sn\ were made on day 1278
(Jakobsen et al. 1991), and we can combine a mass estimate using $\Hb$
at this time with the observed thickness of the ring to infer the 
angular
extent of the ring, $\Delta\theta$, perpendicular to the ring plane, as seen from the 
supernova.  From the [\OIIb\ and [\NIIb\ lines the density and 
temperature of the gas were at this time $n_e \sim 1\EE4\,\cm3$ 
and $T_e \sim 1.7\EE4\rm\,K$ (Wang 1991; Cumming et al. 1996). 
The flux in $\Hb$ on day 1280 was 
$\sim 7.2\EE{-14}\ergcms $ (Wang 1991), and the 
luminosity $2.2\EE{34}\ergs$ (for $\EBV = 0.16$). 
The total mass was therefore  $4.3\EE{-2} (n_e/10^4 \,\cm3)^{-1} 
(T_e/1.7\EE{4} \rm\,K)^{ 0.91} ~\Msun$.
Although characteristic of a different density component, this mass is 
not very different from that near day 400. Jakobsen et al. (1991) estimate
the thickness of the ring in the plane of the sky,  $\Delta R_{\perp}$, to
$\sim 10\%$ of the radius, or $\sim 6\EE{16}\rm\, cm$. Plait et al. (1995)
found a slightly larger value, $\sim 15\%$. Because later images obtained
shortly after day $\sim 2500$ using the COSTAR corrective optics 
(Jakobsen et al. 1994) tend to favor the lower
estimate of $\sim 10\%$, we will use this in our analysis. We note, however,
that the emitting region could have been somewhat more extended at day 1280
without contradicting the observations by Jakobsen et al. (1994).
Assuming a volume filling factor within the ring, $\epsilon$, we get  
\begin{equation}
 \Delta\theta \approx
1.4 {  {}^\circ} \epsilon^{-1} \left( n_e \over 1\EE{4} \rm \,
\cm3\right)^{-2} \left({\Delta R_{\perp}/R \over 0.1}\right)^{-1} \left(T_e
\over 1.7\EE{4} \rm K \right)^{ 0.91} .
\end{equation}
This angular extent corresponds to
a thickness along the line of sight, $R_{\parallel}$,  
\begin{equation}
{\Delta R_{\parallel} \over R } \approx 
0.012 \epsilon^{-1} \left( n_e \over 1\EE{4} \rm \,
\cm3\right)^{-2} \left({\Delta R_{\perp}/R \over 0.1}\right)^{-1} \left(T_e
\over 1.7\EE{4} \rm K \right)^{ 0.91} ,
\end{equation}
around day 1278, i.e., smaller than the radial extent $(\sim 0.1\,R)$, 
unless $\epsilon \ll 1$.  For dynamical reasons we would if anything 
expect the radial extent to be smaller than the lateral.  A solution 
to this problem may be that the filling factor is considerably smaller 
than unity, and/or that $\Delta R_{\perp}/R \ll 0.1$.  The \HST 
[\OIIIb\ images indeed show considerable structure which may indicate 
$\epsilon < 1$ in the ring.  However, it is difficult to see that it 
can be much less than unity {\it if} the ring is resolved in the most 
recent images (Jakobsen et al.  1994), unless the ring consists of 
blobs smaller than the \HST resolution, $\sim$0\farcs1 ($\sim 
7.5\EE{16} \rm~ cm $ at a distance of 50 kpc).  These conclusions 
agree with those of Plait et al.  (1995) and LF96.

\subsection{DENSITY COMPONENTS}

The line emission becomes dominated by gas of decreasing density 
because cooling and recombination time scales are inversely 
proportional to electron density.  Therefore, the flux seen at early 
times is likely to come from gas of considerably higher density than 
at later times (LF96), consistent with the order of magnitude decrease 
in $n_e$ derived in \S 4.4 from high-dispersion \IUE spectra.  This 
agrees with optical results for [\OIIb\ at day 1280 by Wang (1991) who 
finds $n_e \sim 0.7\EE4~\cm3$.  A steady decrease in electron density 
was also found by Plait et al.  (1995) when they analyzed \HST 
[\OIIIb\ images spanning the epoch 1278 -- 2431 days.  The emitting 
mass may therefore vary with time.

Additional information about the density distribution can be inferred 
from the low resolution \IUE observations.  LF96 find it necessary to 
include gas with electron densities down to $\sim 6\EE{3} \cm3$ to 
explain the slow decline of the lines after maximum.  Such a 
distribution of densities in the emitting gas is expected from 
hydrodynamic simulations of the circumstellar structure of \sn\ 
(Blondin \& Lundqvist 1993), which show that the highest densities are 
found close to the ring plane.  They also find that there is a large 
amount of low density gas both in and out of the plane, possibly 
accounting for the gas causing the resonance scattering of \NV\wl1240.
  If the low density gas is ionized up to \NV\ and 
higher, and has a density of $n_e \gtrsim {\rm a~few} \times 10^2 
\cm3$, this may also give rise to a weak, steady line emission, as 
predicted by LF91.  They argued that such emission should be the sign 
of the inner parts of the unshocked red supergiant wind.  In the 
interacting winds scenario (e.g., Blondin \& Lundqvist 1993; Martin \& 
Arnett 1995), the emission could also come from the {\it shocked} red 
supergiant wind (LF96) or perhaps the \HII region interior to the 
ring (Chevalier \& Dwarkadas 1995).  The increasing importance of density 
components with $n_e \lesssim 10^4 \cm3$ after $\sim 1000$ days is 
consistent with the observed decrease in electron density inferred 
from density-sensitive line ratios (\CIIIb, \S 4.4) and the optical 
[\OIIb\ and [\SIIb\ lines (Wampler et al.  1989; Wang 1991; LF96).

There are also some theoretical indications for gas with $n_e \gtrsim 
4\EE{4} \cm3$.  LF96 model three components with different densities 
and find that the bulk of the \CIIIb\ emission at $t \lesssim 400$ days 
originates in their highest density component ($n_e \sim 4\EE{4} 
\cm3$), i.e., close to what we derive in \S 4.4.  It is likely that 
there is a continuous range of densities rather than components with 
a~few discrete densities.  The component with $n_e \sim 4\EE{4} \cm3$ 
in LF96 therefore represents a mean value at the high end of this 
range, indicating the existence of gas with even higher densities.  
The scatter in the \NIVb\ light curve is considerably larger during the 
first 400 days than at later times (Fig.  6c), which may be a further 
sign of components with $n_e \gtrsim 4\EE{4} \cm3$.  Because 
recombination from \NIV\ to \NIII\ is rapid, and collisional ionization 
unimportant, the emissivity of the \NIVb\ line declines rapidly with 
time at a given point in the emitting gas (see Fig.  2 of LF91).  
Therefore, the flux of the \NIVb\ line is not severely smoothed by the 
light echo, and responds rapidly to changes in the density of material  
being swept up by the light echo paraboloid.  When higher density gas 
is encountered, the result is a momentary increase in the 
observed flux, and may thus explain the fluctuations.  Fluctuations 
due to density variations around the ring decrease in importance when 
the full ring is seen and only recombination is important.  The narrow 
spikes in \NV\ and \NIIIb\  between days 350 and 500 (see Fig.  6) are 
also probably due to higher density gas and the `arcsine' shape 
of the light curve, characteristic of a ring geometry (e.g., Felten \& 
Dwek 1991; Lundqvist 1991).

There is also evidence for density structure in the gas after maximum, 
especially in \NIIIb\ and 
\NV\ near day 900 to 1000. These fluctuations indicate that there may be 
regions of high density $\sim 2 - 3$ times further away from the supernova 
than the ring radius. (This flux increase might also arise from the 
outer rings, as discussed in \S 4.1.) Several groups have detected blobs in the optical 
and near-infrared which may be responsible for these fluctuations. 
Allen et al. (1989) detected a concentration of \HeI 1.083 $\mu$m emission, and they 
estimate a gas density of $\sim 10^5 \cm3$. Extended \HeI 1.083 $\mu$m emission 
was also observed by Elias et al.  (1993). Crotts, Kunkel, \& McCarthy (1989) 
and Hanuschik (1991) detected $\Ha$ blobs on days 750 and 950, respectively, 
and Cumming \& Meikle (1993) found evidence for a high density blob with 
short life time on day 1344. Cumming \& Meikle estimate that the gas density 
in 
the blob has to be $(1-2)\EE5 \cm3$ and argue that it may be accounted for by
the high density regions forming off the equatorial plane in the hydrodynamic 
models by Blondin \& Lundqvist (1993).

\subsection{COMPARISON WITH MODELS}

Comparison of the observations with a realistic, self-consistent
photoionization/recombination model of the circumstellar ring is
necessary to test the interpretation of these data. 
Such a model includes solving the coupled, time dependent
equations for temperature and ionization; separating these two aspects
leads to an unphysical model. Knowledge of the emissivity as a function 
of time and position makes it straightforward to incorporate the 
light echo effect. This was first done by Lundqvist \& Fransson (1987), 
and later by Fransson \& Lundqvist (1989) and LF91. In those models it was
assumed that the emission originated in a {\it spherical}
shell of density $2.6 \EE{ 4} \cm3$. For the ionizing burst at
breakout two \sn\ explosion models were used, the 10L model by
Woosley (1988) and the 11E1Y6 model by Shigeyama et al. (1988). 
Qualitatively, both outburst models reproduced the observed 
relative line strengths surprisingly well, considering the lack of 
free parameters. The calculations agree fairly well with the 
rapid increase of the \NIIIb\ and \NV\ lines, 
and the nearly constant flux in the \NIVb\ line during 
the first $\sim 400$ days. 
The steady \HeII\ flux was found to be a result of collisional 
excitation. The model of LF91 based on Shigeyama et al.'s 11E1Y6 
spectrum provided the best agreement with both the behavior of 
the UV lines and the gas temperature, as determined from the 
[\OIIIb\wll$4363/(4959 + 5007)$ ratio. The harder burst spectrum 
by Shigeyama et al. was therefore favored.

A closer examination, however, reveals several deficiencies.  Calculations 
using both the 10L and 11E1Y6 models produced too rapid an increase in 
emission flux, especially the \NV\ line, which rises almost linearly with time 
(Fig. 6a).  As discussed by Lundqvist (1994) and LF96, this may be the 
result of resonance scattering of the \NV\ line, not included in the early 
models.  LF96 also find that the slow decline of the \NV\ line may be a 
result of resonance scattering in gas external to the ring.  The scattering 
gas could extend as much as $\sim 10^{18}$ cm away from the ring plane.  
The fact that the maxima of the \NIIIb, \NIVb, \NV\ and \CIIIb\ lines are all 
observed to occur at nearly the same time poses a more serious problem for 
the LF91 models, which in contrast show the \NIIIb\ and \CIIIb\ lines peaking 
considerably later than \NV.  In addition, the optical [\OIIIb\ lines show a 
persistent emission up to $\sim 1500$ days, in contrast to the early 
models.

There are two important factors not taken into account in the 
first models which probably account for these discrepancies. As 
acknowledged by LF91, the assumption of only one density component 
is artificial, and the observations clearly indicate that
this is not true. A combination of density components, or more 
realistically a continuum of densities, ranging from 
$n_e \sim 6\EE{3} \cm3$ to at least $n_e \sim 4\EE{4} \cm3$, 
gives considerably better agreement, as demonstrated by LF96. 
However, such a distribution
cannot be specified in a unique way as it depends on the detailed structure 
of the ring. The other important assumption in LF91 was that of a 
spherical geometry. A toroidal geometry changes the light curve, 
when integrating over the light echo paraboloid, as discussed by 
Lundqvist (1991), Luo (1991) and Dwek \& Felten (1992).
A generic feature of a ring model compared to a spherical 
shell is the rapid rise in the observed fluxes close to maximum light. 
This basically reflects the amount of emitting gas swept up by the light 
echo paraboloid per unit time. Even more important, 
the larger radial thickness required by a toroidal geometry also
implies that the ionized gas will be optically thick in the EUV continuum. As already
explained, the ionization will then be stratified radially, with several
distinct ionization zones (Lundqvist 1992; LF96). The consequence for
the light curve is to give nearly simultaneous
maxima for the different ions, as is observed. In addition, Ensman \&
Burrows (1992) find that the ionizing spectra used in LF91 are probably 
too soft, as was also pointed out in LF91. Ensman \& Burrows find that the 
SN spectrum at breakout will have a high energy tail due to electron scattering. 
Ensman \& Burrows also note that light travel time effects smooth the 
time variation of the spectrum, as may also be the case if
the progenitor was non-spherical (see Lundqvist 1992 for a more detailed
discussion). However, LF96 find that this effect is unimportant
for the ionization of the ring. 

In summary, the light curves are fully consistent with a circular
ring geometry, both from the rapid rise close to maximum light seen in the
\NIIIb\ and \NV\ lines, the turn-on time of these lines, and the 
near-coincidence of the light curve peaks. A ring geometry also makes it 
easier to understand the low expansion velocity, $\sim 10~\kms$ 
(Luo \& McCray 1991; Blondin \& Lundqvist 1993). The origin of the ring, 
however, remains obscure (e.g., Podsiadlowski 1992 for a review). 
The high N/C and N/O ratios put strong constraints on
the ring implying that it must have been formed after the dredge up phase. 
An interesting proposal by McCray \& Lin (1994) explains the ring as
the result of the gradual evaporation of a circumstellar disk around the 
progenitor. They explain the nitrogen enrichment as a result of gradual spraying 
of the inner boundary of the disk by the red supergiant wind. Any model for
the formation of the ring must also be able to explain the outer rings
observed by Wampler et al. (1990), Wang \& Wampler (1992), Burrows et al. 
(1995) and Plait et al. (1995). The interacting winds model 
(e.g., Luo \& McCray 1991; Blondin \& Lundqvist 1993) may do so, 
as demonstrated qualitatively by Martin \& Arnett (1995), and so may the 
model by Lloyd, O'Brien, \& Kahn (1995). The interacting winds model 
can also account for the continuum observations of Crotts et al. (1995) 
(Martin \& Arnett 1995), and the hot gas responsible for the resonance 
scattering of the \NV\ line (LF96). However, the high density of the outer
rings, a few  $10^3 \cm3$ (estimated by LF96) suggests that
the interacting winds model needs some adjustment. An interesting extension
is that of Chevalier \& Dwarkadas (1995). They assume a lower mass loss
rate during the blue supergiant stage, implying that photoionization by 
the progenitor is important during the shaping of the nebula. 
A test of the models for the formation of the observed structure 
around \sn\ will be possible 
as the supernova ejecta strikes the ring shortly after the year 
$\sim 2000$ (e.g., Masai \& Nomoto 1994; Luo, McCray, \& Slavin 1994;
Chevalier \& Dwarkadas 1995; Borkowski, Blondin \& McCray 1996).

\acknowledgments
We thank Robert Cumming and Jim Felten for comments on the manuscript 
and to Martino Romaniello for measuring the positions of stars 2 and 3 
in WFPC2 images.  G.S. gratefully acknowledges the hospitality and 
support of the Institut d'Astrophysique de Paris during a six-month 
sabbatical leave, during which this paper was completed.  This 
sabbatical was made possible by a NASA Research and Study Fellowship.  
C.F. and P.L. are both supported by the Swedish Natural Sciences 
Research Council and the Swedish National Space Board.  C.F. is also 
supported by the G\"oran Gustafsson Foundation for Research in Natural 
Sciences and Medicine.  This work was supported in part by NASA grant 
NAS5-2487 to Harvard College Observatory.

\clearpage
\begin{deluxetable}{lrccrrrrrr}
\tablewidth{490pt}
\tablecaption{Emission Line Fluxes for SN~1987A Circumstellar Ring}
\tablehead{
\colhead{SWP\tablenotemark{a}}           & \colhead{Date\tablenotemark{b}}      &
\colhead{t$_{\rm expo}$\tablenotemark{c}}  & \colhead{v$_{\rm corr}$\tablenotemark{d}}  &
\multicolumn{6}{c}{Integrated Flux (units: $1\EE{-15} \rm {~erg~cm^{-2}~s^{-1}}$)} \\
\cline{5-10} \\
&\multicolumn{1}{c}{{\scriptsize (days)}} 
&\multicolumn{1}{c}{{\scriptsize (min)}} 
&\multicolumn{1}{c}{{\scriptsize (km/s)}} 
&\multicolumn{1}{c}{{\scriptsize N\,{\tiny V}\,1240}} 
&\multicolumn{1}{c}{{\scriptsize N\,{\tiny IV]}\,1486}} 
&\multicolumn{1}{c}{{\scriptsize He\,{\tiny II}\,1640}} 
&\multicolumn{1}{c}{{\scriptsize O\,{\tiny III]}\,1666}} 
&\multicolumn{1}{c}{{\scriptsize N\,{\tiny III]}\,1750}} 
&\multicolumn{1}{c}{{\scriptsize C\,{\tiny III]}\,1908}} 
}
\startdata        
30440     &   11.3 & 300    & $-257$    & 5 & 13 & 2 & \nodata & 1 & \nodata  \nl
30472     &   14.0 & 180    &   46      &  14 & $-7$ &  14 &  \nodata &   4  &   1  \nl
30512     &   18.3 & 240    &  $-46$    &   2 & $-9$ & $-1$ &  17 & \nodata &   6 \nl
30522     &   19.1 & 181    &  591      &  11 &   4 & $-2$ &   7 &  16  & $-4$ \nl
30547     &   21.3 & 240    &  111      &   6 & \nodata & $-2$ &  17 &  10  &   5 \nl
30592     &   27.3 & 300    & $-125$    &   3 &   7 &  19 & $ -2$ &   6  & $ -5$ \nl
30637     &   33.1 & 185    & $-106$    &   2 &  10 &  18 & $ -5$ &  4  &   5 \nl
30743     &   44.2 & 225    &   53      &   4 &  21 &   4 &   6 &   1  &   2 \nl
30907     &   70.2 & 230    &  164      &   7 &   4 &  26 &  19 &  14  &   7 \nl
30974     &   80.6 & 120    &  303      &  55 &   2 &  49 &  \nodata &  12  &  11 \nl
31000     &   85.2 & 180    &  $-37$    &  69 & 45 &  70 &  12 &  17  &   8 \nl
31040     &   90.9 & 195    &  233      &  64 & 114 &  33 &  35 &  29  &  17 \nl
31064     &   96.3 & 180    &  199      &  84 & 101 &  74 &  12 &  76  &   6 \nl
31125     &  105.8 & 200    &  $-7$     & 105 &  77 &  55 &  27 & 117  &  13 \nl
31132     &  107.8 & 195    &   42      &  92 &  95 &  67 &  42 & 109  &   6 \nl
31154     &  110.8 & 180    & $-154$    & 144 & 107 &  54 &  68 & 113  &  27 \nl
31166     &  113.0 & 180    &  $-83$    & 107 & 108 &  70 &  50 &  88  &  29 \nl
31245     &  121.7 & 260    &    8      &  98 & 155 &  43 &  66 &  90  &  91 \nl
31273     &  128.0 & 240    &  111      &  98 & 131 &  75 &  66 & 191  &  38 \nl
31319     &  138.5 & 240    &    7      & 115 & 163 &  92 &  75 & 170  &  70 \nl
31371     &  147.0 & 240    &  130      & 159 & 117 &  94 &  64 & 215  &  78 \nl
31420     &  154.5 & 240    &  280      & 163 & 141 & 116 &  94 & 207  &  72 \nl
31462     &  160.9 & 240    &  122      & 140 & 139 &  85 & 139 & 190  &  53 \nl
31534     &  168.5 & 220    &  396      & 150 & 117 &  78 &  88 & 265  &  88 \nl
31592     &  177.7 & 240    &  243      & 186 & 136 &  93 &  84 & 183  &  X\tablenotemark{e}  \nl
31651     &  185.4 & 210    & $-282$    & 146 &  92 &  95 & 141 & 222  &  94 \nl
31676     &  188.8 & 240    & $-119$    & 147 & 157 &  60 & 130 & 233  &  X  \nl
31819/{\it 18}  &  198.9 & 240/{\it 90} &  $-32$/{\it 173}  & 163 & 186 &  67 & 108 & 302  & {\it 106} \nl
31893     &  209.6 & 200    & $-104$    & 175 & 181 &  75 &  93 & 268  & 110 \nl
31954     &  220.3 & 200    & $-926$    & 131 & 148 &  74 &  99 & 265  & 101 \nl 
32030/{\it 31}  &  227.7 & 240/{\it 90} & --142/{\it --34}  & 139 & 177 & 110 & 155 & {\it 389} & {\it 101} \nl
32168     &  245.5 & 155    &  520      & \nodata & \nodata & \nodata & \nodata & 343  & 102 \tablebreak

32219/{\it 20}  &  252.6 & 240/{\it 80} & --352/{\it --370} & 176 & 126 &  90 &  85 & {\it 323} & {\it 113} \nl
32314     &  263.4 & 235    &   46      & 215 & 161 &  93 &  94 &  X   &  X   \nl
{\it 32395} & 275.8 &{\it 90} &{\it--500} & \nodata & \nodata & \nodata & \nodata &{\it 321} &{\it 135} \nl
32404     &  276.6 & 240    & $ -98$    & 234 & 221 &  96 &  97 &  X   &  X  \nl
32532     &  296.2 & 240    & $-212$      & 223 & 214 &  85 & 145 &  X   &  X  \nl
32619/{\it 20}  &  305.5 & 240/{\it 80} & --493/{\it --236} & 233 & 202 &  77 & 132 & {\it 344} & {\it 177} \nl
32717     &  324.2 & 190    & $-414$      & 291 & 141 &  \nodata & 111 & 311  & 114 \nl
32797     &  338.4 & 210    & $-189$      & 322 & 170 & 108 & 116 & 411  &  X  \nl
{\it32879} & 349.5 &{\it 75} &{\it--45} & \nodata & \nodata & \nodata & \nodata & {\it 426} &{\it 162} \nl
32911     &  355.1 & 212    & $-170$      & 338 & 198 &  76 &  95 & 414  & 137 \nl
32938     &  360.0 & 180    & $-591$      & 349 & 174 &  69 &  97 & 425  & 114 \nl
33035     &  375.3 & 240    &   85      & 379 & 197 & 116 & 127 &  X   &  X  \nl
33105/{\it 04}  &  388.0 & 240/{\it 70} & 5/{\it --211} & 375 & 211 & 114 & 126 & {\it 482} & {\it 206} \nl
33175/{\it 76}  &  400.3 & 210/{\it 90} &  85/{\it --292} & 464 & 195 &  84 & 149 & 442  & {\it 199} \nl
33280     &  416.0 & 200    &  621      & 498 & 151 & 110 & 172 & 511  & 197 \nl
33331/{\it 32}  &  423.2 & 240/{\it 75} & $135/{\it -186}$ & 383 & 202 &  80 & 173 & {\it 444} & {\it 181} \nl
{\it 33423} & 433.6 &{\it 80} &{\it --40} & \nodata & \nodata & \nodata & \nodata &{\it 338} &{\it 178} \nl
{\it 33492} & 440.6 &{\it 80} &{\it 345} & \nodata & \nodata & \nodata & \nodata &{\it 531} &{\it 176} \nl
33497/{\it 96} & 441.9 & 240/{\it 70} &    3/{\it 292}  & 385 & 152 &  79 & 137 & {\it 404} & {\it 202} \nl
33519/{\it 20} & 445.1 & 240/{\it 80} & --178/{\it 162}  & 418 & 170 &  98 & 131 & {\it 362} & {\it 215} \nl
{\it 33644} & 458.3 &{\it 80} &{\it 73} & \nodata & \nodata & \nodata & \nodata &{\it 322} &{\it 211} \nl
{\it 33725} & 471.4 &{\it 80} & {\it--326} & \nodata & \nodata & \nodata & \nodata &{\it 324} &{\it 143} \nl
33742/{\it 41}  &  473.8 & 235/{\it 80} &  103/{\it 44}   & 348 & 121 &  73 & 119 & {\it 367} & {\it 241} \nl
33799/{\it 800} &  485.0 & 240/{\it 80} &  145/{\it 153}  & 561 & 133 &  85 & 114 & {\it 327} & {\it 178} \nl
33869     &  498.7 & 195    & $-132$      & 343 & 130 &  84 & 147 & 330  & 145 \nl
33937     &  510.9 & 240    &  220      & 303 & 113 &  91 & 106 & 321  &  X  \nl
33967/{\it 66}  &  518.7 & 251/{\it 90} & 172/{\it --236} & 272 & 109 & 112 & 136 & 361  & {\it 144} \nl
34057/{\it 58}  &  532.9 & 240/{\it 80} & 172/{\it --183} & 314 & 118 &  95 &  88 & 266  & {\it 127} \nl
34088     &  538.5 & 200    &  347      & 267 & 141 &  75 &  99 & 294  & 120 \nl
34232     &  567.5 & 200    &  115      & 245 & 105 &  86 &  92 & 270  & 103 \nl
34441     &  593.5 & 150    & $-200$    & 194 & \nodata & \nodata & \nodata & 199  & 100 \nl
34640     &  615.8 & 240    & $-283$     & 203 &  82 &  40 &  98 & 223  & 111 \tablebreak

34751     &  632.7 & 240    & $-400$     & 197 &  61 &  85 &  72 & 222  & 100 \nl
34871     &  650.7 & 163    & $-502$      & 203 & \nodata & \nodata & 104 & 213  &  89 \nl
35126     &  670.7 & 180    & $-691$      & 190 &  63 &  77 & 114 & 167  &  94 \nl
35308     &  686.5 & 220    & $-332$      & 238 &  74 &  65 &  67 & 184  &  99 \nl
35505     &  714.3 & 220    & $-453$      & 187 &  62 &  54 &  87 & 170  &  68 \nl
35686     &  740.3 & 240    &  114      & 176 &  24 &  24 &  73 &  94  &  70 \nl
35822     &  755.1 & 183    &   33      & 182 &  48 &  33 &  74 & 119  & 106 \nl
35940     &  771.2 & 260    &  $-29$      & 152 &  55 &  55 &  50 & 112  &  66 \nl
36171     &  796.2 & 210    &   59      & 135 &  63 &  40 &  \nodata & 123  &  74 \nl
36258     &  809.9 & 222    &  136      & 159 &  21 &  45 &  87 & 131  &  49 \nl
36279     &  813.2 & 260    &  267      & 143 &  68 &  54 &  55 & 129  &  50 \nl
36539     &  844.0 & 260    &   96      & 134 &  25 &  48 &  52 &  99  &  45 \nl
36578     &  853.8 & 310    &    0      & 142 &  35 &  51 &  \nodata & 106  &  54 \nl
36676     &  870.9 & 260    &  141      & 151 &  28 &  47 &  87 & 101  &  33 \nl
36801     &  893.9 & 260    &   93      & 129 &  18 &  44 &  55 & 120  &  58 \nl
36968     &  928.5 & 270    &   89      & 137 &   7 &  28 &  44 &  67  &  43 \nl
37062     &  936.5 & 230    & $-182$      & 119 &   1 &  34 &  70 & 102  &  48 \nl
37088     &  938.8 & 300    &  218      & 115 &   1 &  42 &  54 & 121  &  54 \nl
37236     &  951.5 & 263    &  873      &  89 &   3 &  36 &  55 &  85  &  26 \nl
37424     &  970.7 & 230    & $-359$      & 103 &   5 &  46 &  39 & 108  &  47 \nl
37574     &  992.6 & 300    & $-501$      & 122 &   9 &  51 &  55 &  91  &  52 \nl
37798     & 1021.6 & 300    & $-422$      &  82 &  42 &  44 &  51 &  64  &  62 \nl
37973     & 1049.5 & 285    & $-314$      &  99 &  33 &  69 &  56 &  63  &  39 \nl
38055     & 1061.1 & 280    & $-748$      &  92 & $-9$ &  19 &  60 &  95  &  32 \nl
38172     & 1081.3 & 285    & $-332$      & 101 &   4 &  11 &  42 &  69  &  78 \nl
38307     & 1105.3 & 285    & $-424$      & 135 &  20 &   5 &  43 &  48  & 19 \nl
38336     & 1112.0 & 275    & $-414$      & 115 & $-3$ &  24 &  44 &  50  &  33 \nl
38536     & 1137.2 & 285    &   90      &  89 &   7 &   4 &  57 &  51  &  34 \nl
38866     & 1186.1 & 285    & $-152$      &  80 &  18 &  13 &  22 &  24  &  34 \nl
39300     & 1247.0 & 285    & $-6$      &  50 &  10 &  35 &  55 &  28  &  23 \nl
39757     & 1317.7 & 265    & $-287$      &  90 & $-11$ &   6 &  40 &  35  &  19 \nl
40002     & 1344.7 & 270    & $-550$      &  65 & $-10$ &  53 &  58 &  40  &  10 \tablebreak

40275     & 1380.5 & 280    & $-491$      &  43 &   6 &  23 &  28 &  28  &   9 \nl
40858     & 1449.3 & 280    & $-422$      &   4 &   2 &   3 &  28 &  19  &  \nodata \nl
41179     & 1490.4 & 310    & $-110$      &  17 & $-5$ &   2 &  25 &  16  &  11 \nl
41802     & 1566.0 & 290    &  198      &  20 & $-2$ &  \nodata &  20 &  10  &   3 \nl
42174     & 1623.0 & 270    & $-372$      &  46 &  10 &  32 &  26 &   2  &  15 \nl
43049     & 1719.6 & 340    & $-307$      &  37 & $-10$ &   3 &  22 &   2  &   5 \nl
43393     & 1755.6 & 340    & $-255$      &  21 &   1 &   9 &  10 &  11  & $-4$ \nl
44017     & 1820.4 & 340    & $-678$      &  43 &  10 & $-3$ &  28 &   8  &   7 \nl
44142     & 1840.3 & 360    & $-405$      &  17 &   3 & $-3$ &  24 &   5  & $-1$ \nl
44442     & 1882.2 & 270    & $-577$      &  18 &  19 &   5 &  28 &   9  &   1 \nl
\enddata
\tablenotetext{a}{SWP image number.  On thirteen dates two images of different
exposure times  were taken consecutively to obtain good-quality spectra of all
six lines.  Data entries corresponding to exposures shorter then 100 minutes are given in
italics.}
\tablenotetext{b}{Observation date, measured in days from the IMB neutrino event, 
1987 Feb. 23.316 (Bionta et al. 1987).}
\tablenotetext{c}{Exposure duration in minutes.}
\tablenotetext{d}{Cross-correlation velocity correction (see \S 3.3).}
\tablenotetext{e}{X denotes that the raw spectral data is saturated at this wavelength.}
\end{deluxetable}
\clearpage

\begin{deluxetable}{ccrrrrr}
\tablenum{2}
\tablewidth{350pt}
\tablecaption{High Dispersion C III] and N III] Emission Line Fluxes} \label{tbl-hiobs}
\tablehead{
\colhead{SWP} & \colhead{Date\tablenotemark{a}} & 
\colhead{$t_{expo}\tablenotemark{b}$} & \colhead{$F_{1907}\tablenotemark{c}$} & 
\colhead{$F_{1909}\tablenotemark{c}$} & \colhead{$F_{1749}\tablenotemark{c}$} & 
\colhead{$F_{1752}\tablenotemark{c}$} \\
}
\startdata
   32394   &   275.544  &  450 &  6.277 &   7.085 &   15.84  & 12.63 \nl
   32983   &   367.110  &  540 &  7.513 &   9.318 &   19.64  & 12.72 \nl
   33215   &   408.167  &  880 &  7.924 &   9.832 &   23.76  & 15.12 \nl
   33536   &   448.026  &  900 & 11.973 &   7.996 &   22.17  & 14.53 \nl
   33810   &   487.901  &  850 &  7.139 &   5.311 &   18.96  &  9.84 \nl
   34024   &   528.874  &  952 &  6.790 &   5.560 &   16.25  & 12.52 \nl
   34420   &   590.622  &  760 &  6.229 &   4.578 &   14.07  &  8.43 \nl
   34870   &   650.340  &  612 &  8.094 &   5.099 &   10.57  &  5.72 \nl
   35577   &   726.232  &  925 &  5.634 &   3.530 &   12.69  &  6.15 \nl
   35929   &   771.117  &  630 &  4.303 &   3.468 &  \nodata &  6.90 \nl
   36399   &   834.004  &  880 &  6.821 &   3.311 &    9.96  &  7.59 \nl
   36891   &   917.794  &  945 &  3.234 &   2.897 &    5.48  &  3.10 \nl
   37297   &   960.701  &  975 &  3.236 &   1.567 &    8.05  &  3.44 \nl
   37805   &  1024.533  &  970 &  2.512 &   1.481 &   17.34  &  3.57 \nl
   38230   &  1096.065  &  610 &  2.576 &   1.372 &    5.42  &  2.69 \nl
   39093   &  1208.039  &  965 &  1.623 &   0.907 &    6.66  &  5.04 \nl
   39644   &  1300.799  &  930 &  1.803 &   0.954 &    5.98  &  3.29 \tablebreak
   40268   &  1380.515  & 1010 &  1.589 &   1.071 &  \nodata &  2.81 \nl
   41316   &  1505.059  &  625 &  0.865 &   0.546 &    3.27  &  2.35 \nl
   42370   &  1651.787  &  960 &  0.508 &   0.253 &    3.81  &  1.73 \nl
\enddata
\tablenotetext{a}{Exposure mid-point measured in days since 1987 Feb. 54.316 UT}
\tablenotetext{b}{Exposure time duration in minutes}
\tablenotetext{c}{Integrated line flux in units of 
$1\times10^{-14}\ergcms$}
\end{deluxetable}
\clearpage

\begin{deluxetable}{lcrccr}
\tablenum{3}
\tablewidth{350pt}
\tablecaption{Extinction Corrections} \label{tbl-3}
\tablehead{
\colhead{Line} & \colhead{\EBV=0.16\tablenotemark{a}} & 
\colhead{$L_{35}\tablenotemark{b}$} & \colhead{~~~~~}   & 
\colhead{\EBV=0.20\tablenotemark{a}}   & 
\colhead{$L_{35}\tablenotemark{b}$} \\
}
\startdata
 \NV\wl1240         &  5.75 & 1.73 & 	 & 9.32 & 2.80  \nl
 \NIVb\wl1486       &  4.18 & 1.25 &	 & 6.19 & 1.86  \nl
 \HeII\wl1640       &  3.76 & 1.13 &	 & 5.34 & 1.60  \nl
 \OIIIb\wl1666      &  3.68 & 1.10 &	 & 5.20 & 1.56  \nl
 \NIIIb\wl1750      &  3.61 & 1.08 &	 & 5.08 & 1.52  \nl
 \CIIIb\wl1908      &  3.59 & 1.08 &	 & 5.01 & 1.50  \nl
\enddata
\tablenotetext{a}{Extinction correction for observed flux}
\tablenotetext{b}{Line luminosity in $10^{35}\ergs$, per unit 
flux of $1\EE{-13}\ergcms$, assuming $d=50\rm\,kpc$ and the indicated extinction.}
\end{deluxetable}

\begin{deluxetable}{crccrc}
\tablenum{4}
\tablewidth{260pt}
\tablecaption{CNO Abundance Ratios}
\tablehead{
\colhead{$T_e/10^4\rm\,K$} & \colhead{N/C} & \colhead{Error} & \colhead{~~~~} & 
\colhead{N/O} & \colhead{Error} \\
}
\startdata
 1.0                & 15.7  &  2.9 & ~    & 1.4   & 0.3   \nl
 2.0                &  9.4  &  1.7 & ~    & 1.5   & 0.4   \nl
 5.0                &  6.1  &  1.1 & ~    & 1.7   & 0.5   \nl
 10.0               &  4.7  &  0.9 & ~    & 1.8   & 0.5   \nl
\enddata
\end{deluxetable}


\clearpage


\begin{thebibliography}{}

\bibitem {} Aggarwal, K. M., Berrington, K. A., Eissner, W., \& 
     Kingston, A. E. 1986, Recommended Data from the Daresbury Atomic Data 
     Workshop 1985 (Belfast: Daresbury Atomic Data Bank)
\bibitem {} Allard, N., Artru, M.-C., Lanz, T., \& Le Dourneuf, M. 1990, 
       \aaps, 84, 563
\bibitem {} Allen, D. A., Meikle, W. P. S., \& Spyromilio, J. 1989, Nature,
     342, 403 
\bibitem {} Ball, L., Campbell-Wilson, D., Crawford, D. F., \& Turtle, A. 
     J., 1995, \apj, 453, 864
\bibitem {} Berrington, K. A. 1985, J. Phys. B, 18, L395
\bibitem {} Beuermann, K., Brandt, S., \& Pietsch, W. 1994, \aap, 281, L45
\bibitem {} Bionta, R. M., et al. 1987, \prl, 58, 1494
\bibitem {} Blinnikov, S. I., \& Nadyozhin, D. K. 1991, in Supernovae, ed. S. E.
     Woosley (New York: Springer), 213
\bibitem {} Blondin, J. M., \& Lundqvist, P. 1993, \apj, 405, 337
\bibitem {} Blum, R. D., \& Pradhan, A. K. 1992, \apjs, 80, 425
\bibitem {} Bohlin, R. C., \& Grillmair, C. J. 1986, \apjs, 66, 209
\bibitem {} Bohlin, R. C., \& Holm, A. V. 1980, NASA IUE Newsletter No. 10, 37
\bibitem {} Borkowski, K., Blondin, J.~M., \& McCray, R. 1996, preprint
\bibitem {} Burrows, C. J., et al. 1995, \apj, 454, 680
\bibitem {} Cassatella, A., et al. 1987, \aap, 177, L29
\bibitem {} Cassatella, A., et al. 1994, \aap, 281, 594
\bibitem {} Chevalier, R. A. 1988, Nature, 332, 514 
\bibitem {} Chevalier, R. A. 1992, Nature, 355, 691
\bibitem {} Chevalier, R. A., \& Dwarkadas, V. V. 1995, \apj, 452, L45
\bibitem {} Crotts, A. P. S., \& Heathcote, S.R. 1991, Nature, 350, 683   
\bibitem {} Crotts, A. P. S., \& Kunkel, W. E. 1991, \apj, 366, L73 
\bibitem {} Crotts, A. P. S., Kunkel, W. E., \& Heathcote, S. R. 1995, \apj,
     438, 724
\bibitem {} Crotts, A. P. S., Kunkel, W. E., \& McCarthy, P. J. 1989, \apj,
     347, L61
\bibitem {} Cumming, R. J. 1994, Ph.D. thesis, Imperial College, London
\bibitem {} Cumming, R. J., Lundqvist, P., Meikle, W. P. S., Spyromilio, J.,
     \& Allen, D. A. 1996, in preparation
\bibitem {} Cumming, R. J., \& Meikle, W. P. S. 1993, \mnras, 262, 689
\bibitem {} Duffy, P., Ball, L., \& Kirk, J. G. 1995, \apj, 447, 364
\bibitem {} Dwek, E., \& Felten, J. E. 1992, \apj, 387, 551
\bibitem {} Elias, J. H., Phillips, M. M., Suntzeff, N. B., Walker, A. R.,
     Gregory, B., \& Depoy, D. L. 1993, in Proc. Massive Stars: Their Lives 
     in the Interstellar Medium, ed. J. P. Cassinelli \& E. B. Churchwell
     (San Francisco: Astr. Soc. Pacific), 408
\bibitem {} Ensman, L., \& Burrows, A. 1992, \apj,  393, 742  
\bibitem {} Felten, J. E., \& Dwek, E. 1991, in \sn\ and Other
     Supernovae I.J. Danziger \& K. Kj\"ar (Munich: ESO), 569 
\bibitem {} Fitzpatrick, E. L. 1985, \apj, 299, 219
\bibitem {} Fitzpatrick, E. L., \& Walborn, N. R. 1990, AJ, 99, 1483
\bibitem {} Fleming, J., Hibbert, A., \& Stafford, R. P. 1994, Phys.Scr. 49, 316
\bibitem {} Fransson, C. 1994, in Les Houches, Session LIV
     1990, eds. J. Audouze, S. Bludman, R. Mochkovitch, \& J. Zinn-Justin,
     Elsevier Science B.V., 677 
\bibitem {} Fransson, C., Cassatella, A., Gilmozzi, R., Kirshner, R. P.,
     Panagia, N., Sonneborn, G., \& Wamsteker, W. 1989, \apj, 336, 429 
     (Paper I) 
\bibitem {} Fransson, C., \& Lundqvist, P. 1989, \apj,  341, L59 
\bibitem {} Fransson, C., \& Sonneborn, G. 1994, in Frontiers of Space and
     Ground-based Astronomy, eds. W. Wamsteker, M.~S. Longair, \& Y. Kondo
     (Kluwer: Dordrecht), 249
\bibitem {} Garhart, M. P. 1992a, NASA IUE Newsletter No. 48, 80
\bibitem {} Garhart, M. P. 1992b, NASA IUE Newsletter No. 48, 98
\bibitem {} Gilmozzi, R., et al. 1987, Nature, 328, 318
\bibitem {} Gorenstein, P., Hughes, J. P., \& Tucker, W. H. 1994, \apj, 
     420, L25
\bibitem {} Gould, A. 1994, \apj, 425, 51
\bibitem {} Gould, A. 1995, \apj, 452, 189
\bibitem {} Hanuschik, R. W. 1991, \aap, 217, L21
\bibitem {} Hasinger, G., Aschenbach, B., \& Tr\"umper, J. 1996, \aap, 
     312, L9
\bibitem {} Hauschildt, P. H., Starrfield, S., Austin, S., Wagner, R. M.,
     Shore, S., \& Sonneborn, G. 1994, \apj, 422, 831
\bibitem {} Jakobsen, P., et al. 1991, \apj, 369, L63 
\bibitem {} Jakobsen, P., Jedrzejewski, R., Macchetto, F., \& Panagia,
     N. 1994, \apj, 435, L47
\bibitem {} Kahn, I., \& Duerbeck, H. W. 1991, in \sn\ and Other
     Supernovae I.J. Danziger \& K. Kj\"ar (Munich: ESO), 251 
\bibitem {} Keenan, F. P., Berrington, K. A., Burke, P. G., Dufton, P. L., \&
Kingston, A. E. 1986, Phys. Scripta, 34, 216
\bibitem {} Keenan, F. P., Feibelman, W., \& Berrington, K. A. 1992, \apj, 389, 443
\bibitem {} Kinney, A. L., Bohlin, R. C., \& Neill, J. D. 1991, \pasp, 103, 694
\bibitem {} Kirshner, R. P., Sonneborn, G., Crenshaw, D. M., \& Nassiopoulos,
     G. E. 1987, \apj, 320, 602
\bibitem {} Li, H., \& McCray, R., 1996, \apj, 456, 370
\bibitem {} Lloyd, H. M., O'Brien, T. J., \& Kahn, F. D. 1995, \mnras, 273, L19
\bibitem {} Lucy, L. B. 1987, \aap, 182, L31
\bibitem {} Lundqvist, P. 1991, in \sn\ and Other
     Supernovae, eds. I.J. Danziger \& K. Kj\"ar (Munich: ESO), 607 
\bibitem {} Lundqvist, P. 1992, \pasp, 104, 787
\bibitem {} Lundqvist, P. 1994, in Circumstellar Media in the Late Stages of
     Stellar Evolution, ed. R. E. S. Clegg, W. P. S. Meikle, \& I. R. Stevens
     (Cambridge: Cambridge Univ. Press), 213
\bibitem {} Lundqvist, P., \& Fransson, C. 1987, in Proc. ESO Workshop on 
     the \sn, ed. I.J. Danziger \& K. Kj\"ar, (Munich: ESO), 495
\bibitem {} Lundqvist, P., \& Fransson, C. 1991, \apj,  380, 575 (LF91)
\bibitem {} Lundqvist, P., \& Fransson, C. 1996, \apj, 464, 924 (LF96)
\bibitem {} Luo, D. 1991, Ph.D. thesis, Univ. Colorado
\bibitem {} Luo, D., \& McCray, R. 1991, \apj,  379, 659
\bibitem {} Luo, D., McCray, R., \& Slavin, J. 1994, \apj, 430, 264
\bibitem {} Martin, C. M., \& Arnett, D. 1995, \apj, 447, 378
\bibitem {} Masai, K., \& Nomoto, K., 1994, \apj, 424, 924
\bibitem {} McCray, R., \& Lin, D. N. C. 1994, Nature, 369, 378
\bibitem {} Meikle, W. P. S., Allen, D. A., Spyromilio, J., Cumming, R. J.,
     Varani, G.-F., \& Mobasher, B. 1991, in Proc. ESO/EIPC Workshop, 
     \sn\ and other Supernovae, ed. I. J. Danziger \& K. Kj\"ar 
     (Munich: ESO), 265
\bibitem {} Mendoza, C., \& Zeippen, C. J. 1983, \mnras, 202, 981
\bibitem {} Menzies, J. W. 1991, in Proc. ESO/EIPC Workshop, \sn\ and Other
     Supernovae I.J. Danziger \& K. Kj\"ar (Munich: ESO), 209 
\bibitem {} Munoz Piero, J. R. 1985, ESA IUE Newsletter, No. 23, 58; 
     reprinted in NASA IUE Newsletter, No. 27, 27
\bibitem {} Nussbaumer, H., \& Schild, H. 1979 \aap, 75, L17
\bibitem {} Nussbaumer, H., \& Storey, P. J. 1978, \aap 64, 139 
\bibitem {} Nussbaumer, H., \& Storey, P. J. 1979 \aap, 71, L5
\bibitem {} Panagia, N., Gilmozzi, R., Macchetto, F., Adorf, H.-M., \& 
     Kirshner, R. P. 1991, \apj, 380, L23 (Erratum: \apj, 386, L31)
\bibitem {} Plait, P. C., Lundqvist, P., Chevalier, R. A., \& Kirshner, R. P. 
     1995, \apj, 439, 730
\bibitem {} Podsiadlowski, P. 1992, \pasp, 104, 717
\bibitem {} Pun, C. S. J., et al. 1995, \apjs, 99, 223
\bibitem {} Saio, H., Nomoto, K., \& Kato, M. 1988, Nature, 334, 508   
\bibitem {} Sanz Fern\'andez de C\'ordoba, L. 1993, \aap, 276, 103
\bibitem {} Savage, B. D., \& Mathis, J. 1979, \araa, 17, 73
\bibitem {} Scuderi, S., Panagia, N., Gilmozzi, R., Challis, P. M., \&
     Kirshner, R. P. 1996, \apj, 465, 956
\bibitem {} Seaton, M. J. 1987, J. Phys. B. 20, 6431
\bibitem {} Shigeyama, T., \& Nomoto, K. 1990, \apj, 360, 242
\bibitem {} Shigeyama, T., Nomoto, K., \& Hashimoto, M. 1988, \aap, 196, 141  
\bibitem {} Sonneborn, G., 1991, in Supernovae, Proc. Tenth Santa Cruz Summer
     Workshop, ed. S.E. Woosley (New York: Springer), 125  
\bibitem {} Sonneborn, G., Altner, B., \& Kirshner, R. P. 1987, \apj, 323, L35
\bibitem {} Sonneborn, G., et al. 1990, in Evolution in
     Astrophysics: IUE Astronomy in the Era of New Space Missions, ESA SP-310 
\bibitem {} Sonneborn, G., et al. 1988, IAU Circ. No. 4685
\bibitem {} Staveley-Smith, L., et al. 1992, Nature, 355, 147
\bibitem {} Staveley-Smith, L., et al. 1993, Nature, 366, 136
\bibitem {} Thompson, R. W. 1988, NASA IUE Newsletter No. 35, 133
\bibitem {} Turnrose, B. E., \& Thompson, R. W. 1984, IUE Image Processing
     Information Manual, Version 2.0 (CSC/TM-84/6058)
\bibitem {} Turtle, A. J., et al. 1987, Nature, 327, 38
\bibitem {} Walborn, N. R., Lasker, B. M., Laidler, V. G., \& Chu, Y.-H.
     1987, \apj, 321, L41
\bibitem {} Walborn, N. R., Phillips, M. M., Walker, A. R., \& Elias, 
     J. H. 1993, \pasp, 105, 1240
\bibitem {} Walker, A. R., \& Suntzeff, N. B. 1990, \pasp, 102, 648
\bibitem {} Wampler, E. J., \& Richichi, A. 1989, \aap, 217, 31 
\bibitem {} Wampler, E. J., Richichi, A., \& Baade, D. 1989, in IAU Colloquium 
     120, Structures and Dynamics of the Interstellar Medium, eds. G. 
     Tenorio-Tagle, M. Moles \& J. Melnick (Berlin: Springer), 180  
\bibitem {} Wampler, E. J., Wang, L., Baade, D., Banse, K., D'Odorico, S.,
     Gouiffes, C., \& Tarenghi, M. 1990, \apj, 362, L13 
\bibitem {} Wang, L. 1991, \aap, 246, L69
\bibitem {} Wang, L., \& Wampler, E. J. 1992, \aap, 262, L9
\bibitem {} Weiss, A. 1991, in \sn\ and Other Supernovae, eds. I.J. 
     Danziger \& K. Kj\"ar (Munich: ESO), 3 
\bibitem {} West, R. M., Lauberts, A., Jorgensen, H., \& Schuster, H.-E.
     1987, \aap, 177, L1 
\bibitem {} Wood, P. R., \& Faulkner, D. J. 1987, IAU Circ. No. 4739
\bibitem {} Woosley, S. E. 1988, \apj, 330, 218 
\end{thebibliography}
\end{document}